\documentclass[usenatbib,useAMS]{mnras}

\usepackage{amsfonts}
\usepackage{amsmath}
\usepackage{wasysym}
\usepackage{graphicx}
\usepackage{longtable}
\usepackage{lscape}
 
\begin{document}
	
\interfootnotelinepenalty=10000

\label{firstpage}

\title[Comparing GC systems in Fornax and Virgo]{Differences between the globular cluster systems of the Virgo and Fornax Galaxy Clusters}

\author[Dabringhausen, Fellhauer \& Mieske] {
J{\"o}rg Dabringhausen$^{1}$ \thanks{E-mail: joerg@sirrah.troja.mff.cuni.cz},
Michael Fellhauer$^{2}$ \thanks{mfellhauer@astro-udec.cl} and,
Steffen Mieske$^{3}$ \thanks{smieske@eso.org} \\
$^{1}$ Astronomicky ustav, Universita Karlova, V Holesovickach 2, 180 00 Prague, Czech Republic \\
$^{2}$ Departamento de Astronom{\'i}a, Universidad de Concepci{\'o}n, Avenida Esteban Iturra s/n Casilla 160-C, Concepci{\'o}n, Chile \\
$^{3}$ European Southern Observatory, Alonso de Cordova 3107, Vitacura, Santiago, Chile}

\pagerange{\pageref{firstpage}--\pageref{lastpage}} \pubyear{2021}

\maketitle

\begin{abstract}
It is well known that Globular cluster systems are different among galaxies. Here we test to which degree these differences remain on the scale of galaxy clusters by comparing the globular clusters (GCs) in optical surveys of the Virgo galaxy cluster (ACSVCS) and the Fornax galaxy cluster (ACSFCS) in Kolmogorov-Smirnoff Tests. Both surveys were obtained with the Advanced Camera for Surveys (ACS) on board the Hubble Space Telescope, and contain thousands of GCs in dozens of galaxies each. Also well over 100 point sources in the Chandra X-ray Observatory source catalogue were attributed to the GCs in both optical catalogues, and interpreted as low-mass X-ray binaries (LMXBs). Thus, the optical and X-ray data are as uniform as possible. Our main findings are as follows: (1) The spread in luminosities and half-light radii is larger in the ACSVCS than in the ACSFCS. (2) The ratio between the half-light radii for the F475W-passband and the F850LP-passband is on average smaller in the ACSVCS. (3) The distribution of the LMXBs with the luminosity of the GCs is different between both surveys. These findings are significant. The first finding could be a consequence of a wider spread in the distances of the GCs in the ACSVCS, but the others must have internal reasons in the GCs. Thus, the GC systems are also different on a galaxy cluster scale.
\end{abstract}


\section{Introduction}
\label{sec:introduction}

\subsection{The differences between globular clusters systems}
\label{sec:GCs}

The diversity of systems of globular clusters (GCs) around galaxies is comparable with the diversity of the galaxies themselves.

For instance, let $S_N = N_{GC}\times 10 ^ {0.4(M_V+15)}$ be the specific frequency of GCs, where $N_{GC}$ is the number of GCs in the GC-system and $M_V$ is the $V$-band luminosity of its galaxy \citep{HarrisvdBergh1981}. $S_N$ is therefore the number of globular clusters (GCs) per unit luminosity of a galaxy. \citet{Goudfrooij2003} found that the bulges of the five spiral galaxies they observed with Hubble types of b or later, or equivalently bulge-to-total ratios $\le 0.3$, were all consistent with $S_N=0.55 \pm 0.25$. \citet{Chandar2004} essentially confirmed this with additional late-type galaxies. However, the $S_N$ of luminous, and consequently massive Ellipticals is higher than for the bulges of Spirals ($S_N=1.9\pm 0.6$; \citealt{Goudfrooij2003}). For low-mass Ellipticals, $S_N$ is similarly high, but there is a ´valley' in $S_N$ for intermediate masses \citep{Mieske2014}.

Besides different values for $S_N$, the colours of the GCs are often also bimodal, see \citet{ZepfAshman1993} and \citet{Ostrov1993} for the first discussions of this phenomenon. This is primarily seen for ellipticals in dense environments, but see \citet{Spitler2006} for a massive Sa-galaxy in a loose group which also has a bimodal GC-system. In general, it could be that the bi-modality of the GC-systems in Spirals is often not discovered, because of their on average smaller numbers than in Ellipticals (see e.g. \citealt{BrodieStrader2006}).

Characterizing the colour bi-modality more in detail, \citet{Larsen2001} found for the 17 galaxies they discuss that the peak of the blue, metal-poor subpopulation is on average at a $(V-I)_0$-colour of 0.95, and the peak of the red, metal-rich population is on average at a $(V-I)_0$-colour of 1.18. Fig.~4 in \citet{Larsen2001} also indicates that there are large differences in the number ratios between the red and blue populations. For instance, NGC 4406 has a much stronger blue population than NGC1399. 

There are dependencies between the average colours of the red and the blue subpopulation and the luminosity of their galaxy. \citet{Forbes1997} found a dependency for the red subpopulation, and \citet{Strader2004} found a much weaker dependency for the blue subpopulation. Both are such that they are redder for brighter galaxies. Also a dependency between colour of individual GCs of the blue subpopulation and their luminosities was discovered for many galaxies (the ´blue tilt'), but for others not (e.g. \citealt{Strader2006}).

Connected to the finding of a red and a blue subpopulation is the finding of the different average radii of their globular clusters. Metal-poor (blue) globular clusters are about 20 per cent more extended than metal-rich (red) globular clusters (e.g. \citealt{Kundu1998,Kundu1999,Larsen2001,Kundu2001}).

But the average extensions of GCs may also change among the GC-systems of various galaxies in general. If for no other reason, this should be a consequence of the different radial profiles for the GC-distribution around the galaxies. For instance, a slope of -2 to -2.5 with distance to the centre is typical for the radial distributions of low-mass ellipticals, while for the most massive ellipticals, it is -1.5 or even lower (\citealt{BrodieStrader2006}; see also \citealt{Harris1986}). With increasing distance from the centre of the galaxy, also the extensions of the GCs increase (see for example \citealt{vandenBergh1991} and fig.~8 in \citealt{McLaughlin2000} for the Milky Way). Thus, also the ratio of compact versus extended GCs changes among the galaxies. As an example, consider the Milky Way, which has an average extension of 3.9 pc for its GCs \citep{McLaughlin2000}, compared to the Virgo galaxy cluster, which has an average extension of $2.7 \pm 0.35$ pc for its known GCs \citep{Jordan2005}.

\subsection{The peculiarities of UCDs}
\label{sec:UCDs}

At least some of the ultra-compact dwarf galaxies (UCDs) are often considered to be the massive end of the globular cluster luminosity function (e.g. \citealt{Mieske2002,Mieske2012}). However, the UCDs do show some peculiarities that set them apart from the smaller GCs. For instance, they are on average more extended. They also have on average larger mass-to-light ratios (e.g. \citealt{Hasegan2005,Dabringhausen2008}). This has lead some authors to argue that UCDs are at the very least a mixture of extremely massive GCs and other objects (e.g. \citealt{Hasegan2005,Pfeffer2014}), while others tried to explain the on average different properties of UCDs with the different dynamics of objects above a mass of $\approx 10^6 \, {\rm M}_{\odot}$ (e.g. \citealt{Dabringhausen2009,Marks2012}).

In any case, one of puzzling properties of UCDs is that the UCDs in the Virgo galaxy cluster seem different from those in the Fornax galaxy cluster. While \citet{Dabringhausen2008} found that the mass-to-light ratios of UCDs in general cannot be explained with the same stellar initial mass function (IMF) like in open clusters in the Milky Way (see e.g. \citealt{Kroupa2001}), \citet{Chilingarian2008} found for UCDs in the FGC that their mass-to-light ratios are {\it consistent} with the IMF in the Milky Way. Indeed, different average mass-to-light ratios were stated for UCDs in the FGC and the VGC in \citet{Mieske2008}, with the mass-to-light ratios for UCDs in the VGC being on average about 50 per cent higher than in the FGC.

\citet{Dabringhausen2012} looked at the low-mass X-ray binaries (LMXBs) in GCs and UCDs. LMXBs are binaries which consist of a neutron star (NS) or a black hole (BH) and an evolving low-mass star. The low-mass star thereby expands and some of its matter spills over to the NS or BH. This matter emits X-rays, as it is heated up in the accretion process. The X-rays in turn make the NS or the BH observable \citep{Pringle1972}. Thus, an overabundance of LMXBs could indicate an overabundance of stellar remnants. Indeed \citet{Dabringhausen2012} claimed that if there are as many LMXBs as expected in the ´classical' GC-range, then there are too many in the UCD-range. They fixed this mismatch with a variable IMF. This IMF becomes the more abundant in massive stars (that is, top-heavy), the higher the total mass of its UCD is. On the other hand, \citet{Phillipps2013} found that the UCDs in the FGC are remarkably poor in LMXBs, if compared to the LMXB-content of other stellar systems in Fornax. In other words, the study of \citet{Phillipps2013} does not support a top-heavy IMF, in contrast to \citet{Dabringhausen2012}.

This could be interpreted as that the UCDs in the FGC contain fewer LMXBs than the UCDs in the VGC, and thus fewer NSs and BHs. This would be consistent with the UCDs in the FGC also having mass-to-light ratios that are on average a bit lower than the ones of the Virgo-UCDs. But if the UCDs really are the ´tip of the iceberg' of the GC-population in the VGC and the FGC, are those differences then also seen in lower-mass GCs? Or in other words, are the GC-systems of the VGC and the FGC different, either because the number of GCs is still not big enough or because of the differences in the galaxy clusters themselves? The latter could for instance be because the FGC is more compact but less massive than the VGC \citep{Blakeslee2009}. Or do the GC-systems of the whole galaxy clusters approach each other in quantities like average radii and metallicities? This would then be a consequence of the galaxy clusters being a mixture of galaxies of all types and sizes, an thus also with diverse GC systems.

To bring some light into these questions, we compare the GC-systems of the FGC to the ones of the VGC. Thus, in contrast to earlier studies, we are not interested in the GC systems of individual galaxies, but in the GC system of the whole VGC combined against the one of the FGC combined. The particular quantities that we look at are the luminosities and consequently the masses, the radii and the colours of the GCs. The latter is interpreted as a measure of their metallicities. We also look at the probability of the GCs to have an LMXB. This quantity also depends on the masses, radii and metallicities of the GCs.

In Section~\ref{sec:data} the optical data, the X-ray data, and how they are combined for the GCs in the VGC and in the FGC are discussed. Section~\ref{sec:probability} is about the probability to find a LMXB in a GC, and the computer programme to model this probability. Section~\ref{sec:results} is about the results obtained in this paper. Section~\ref{sec:Dabringhausen2012} puts these results into the context of the results obtained in \citet{Dabringhausen2012}. Section~\ref{sec:summary} summarizes the paper.

\section{Data}
\label{sec:data}

\subsection{Optical data}
\label{sec:optical}

\begin{figure*}
\centering
\includegraphics[scale=0.95]{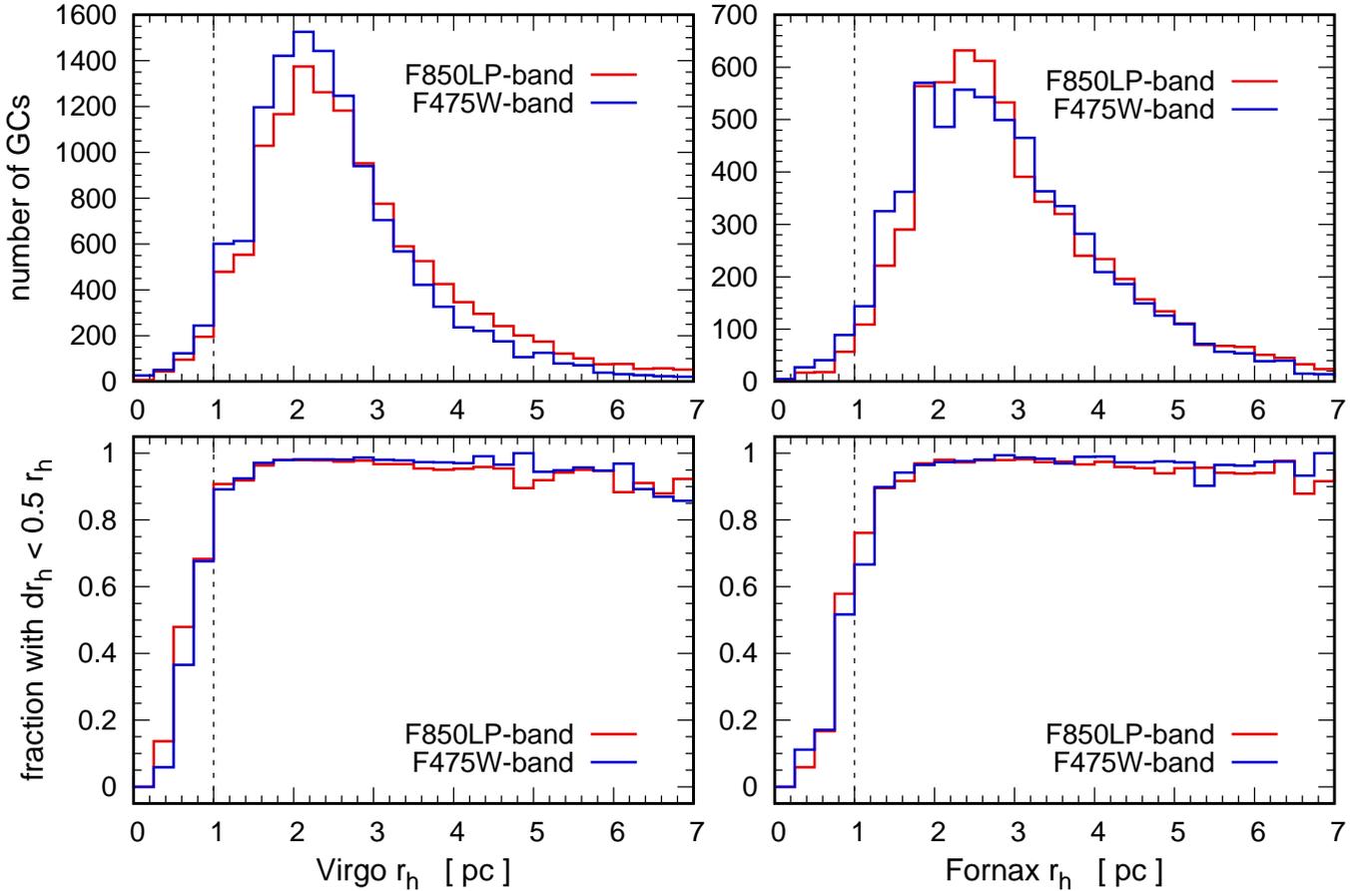}
\caption{\label{fig:cut-radius}
The radii of the GCs in the F475W-band (blue curves) and the F850LP-band (red curves). The top panels shows the distribution of the $r_{\rm h}$ of the GCs. The width of the bins is $0.25\,$pc. The bottom panels shows which fraction of the GC has errors in $r_{\rm h}$ smaller than half the value itself, and therefore pass the appropriate selection criterion. The black dashed lines mark $r_{\rm h} = 1\,$pc. By having $r_{\rm h}>1\,$pc, the GCs pass another selection criterion. The left panels are for the VGC and the right panels are for the FGC.}
\end{figure*}

We take the advanced camera surveys (ACS) on board the Hubble Space Telescope (HST) for the globular cluster calalogues. For the VGC, the raw data was taken in the ACS Virgo cluster survey (ACSVCS, \citealt{Cote2004}), and for the FGC in the ACS Fornax cluster survey (ACSFCS, \citealt{Jordan2007}). Catalogues for GC candidates in these surveys were published in \citet{Jordan2009} for the VGC and in \citet{Jordan2015} for the FGC. These catalogues contain 12763 likely GCs out of 20375 objects for the VGC and 6275 likely GCs out of 9136 objects for the FGC. They are distributed over 100 galaxies in the VGC and 44 galaxies in the FGC. 

However, their distribution is most uneven. For instance, for the VGC, there are 1745 GC-candidates in M87, but only 11 GC-candidates in VCC33. This is because of the different sizes of the GC systems of different galaxies and can therefore not be solved by, for instance, observing VCC33 in more detail.

The luminosities in the F475W-passband ($\approx$ Sloan $g$), the F850LP-passband ($\approx$ Sloan $z$) and half-light radii in these passbands for the GC-candidates are of particular interest for this paper. Note that the catalogues offer two options for the luminosity in each passband: one measured with \textsc{SExtractor} \citep{Bertin1996} and one modeled based on the \textsc{SExtractor} data using \textsc{Kingphot} \citep{Jordan2005}. We take the luminosities based on \textsc{Kingphot} here, because also the radii are based on those data.

We cleanse the two catalogues from data that we consider to be the most imprecise by making four cuts to the catalogues. This is especially necessary for the estimates for the GC-candidates to contain an LMXB. For them, it is not only important whether it is probable for an object to be a GC in the VGC, and FGC, respectively, but also whether the available data is good enough to make comparatively sound predictions. 

The first of these cuts is aimed at objects which are so compact, that they are at the verge of resolution for the HST. We thus remove GC-candidates from the catalogues which have radii $r_{\rm h} < 1 \,$pc in the F475W-passband, and in the F850LP-passband, respectively. The single catalog of GC-candidates of the VGC, and the FGC, respectively, thereby becomes two nearly identical catalogues, one each for every passband. They are different because $r_{\rm h} < 1\,$pc in the F475W-passband does not imply the same for the F850LP-passband, and vice versa.

Naturally, this cut also depend on the assumed distance estimates, for which we take 16.5 Mpc for GC-candidates in the VGC \citep{Mei2007}, and 20.0 Mpc for GC-candidates in the FGC \citep{Blakeslee2009}. However, these estimates are by no means certain, and they neglect that the galaxy clusters have also some depth on the sky (see Section~\ref{sec:directR}). 

The second cut is aimed at GC-candidates which have errors on $r_{\rm h}$ of more than half of $r_{\rm h}$ itself. This cut is made with respect to the theoretical probability of the GC-candidates to contain a low-mass X-ray binary (LMXB), for which we assume $P \propto r_{\rm h}^{-2.5}$ (e.g. \citealt{Sivakoff2007,Dabringhausen2012} and Section~\ref{sec:LMXBtheory}). Errors on $r_{\rm h}$ of more than half of $r_{\rm h}$ itself allow already for a maximum uncertainty of more than 10 in the theoretical probability to have an LMXB in a GC. 

The third and the forth cuts concern the errors in the two passbands which were measured, which have both to be smaller than 0.1 mag. This is because precise magnitudes in both passbands are important for the colours, which are the basis for the estimates of metallicities.

Fig.~\ref{fig:cut-radius} shows the distributions of the radii of the GCs in the F475W-passband (blue curves) and the F850LP-passband (red curves) in its top panels. It can be seen that the fraction of GCs with $r_{\rm h}< 1\,$pc is indeed small.

The lower panels of Fig~\ref{fig:cut-radius} shows which fraction of GCs per bin of $0.25\,$pc fulfills the criterion that the error to $r_{\rm h}$ is less than half the value of $r_{\rm h}$ itself. It seems that additionally requesting that the GCs have $r_{\rm h}<1\,$pc is needless, since the errors in $r_{\rm h}$ become large not only in relative numbers, but also in absolute numbers when the resolution limit in $r_{\rm h}$ is approached.

Nevertheless, the GCs in the FGC and the VGC become more comparable to each other by setting the the minimum $r_{\rm h}$ to the same value of $1\,$pc. Otherwise, this lower limit would have been set by the resolution limit of the HST, which is always the same. Thus, since the VGC is closer than the FGC, there could be compact GCs which are only resolvable at the distance of the VGC, and thus considered there, but are left out as unresolvable at the distance FGC.

Also, \citet{vandenBergh2010} lists the $r_{\rm h}$ of hundreds of GCs in the Milky Way and M31, which are easier to observe because they are closer. However, just two of them have $r_{\rm h}<1\,$pc, and each of them just barely. This casts serious doubts on those objects in the catalogues from \citet{Jordan2009}, and \citet{Jordan2015}, respectively, which undercut this limit in $r_{\rm h}$ severely. It seems more likely that those values are at the resolution limit of the HST, and thus the values in $r_{\rm h}$ are not trustworthy. If they were left in the samples anyway, these GCs would have extremely high theoretical probabilities to have an LMXB, which may be at odds with the observations. Thus, they are not considered here.

We thus get four samples in the optical bands: 10806 GC-candidates in the VGC in the F475W-passband, 10774 GC-candidates in the VGC in the F850LP-passband, 5403 GC-candidates in the FGC in the F475W-passband and 5354 GC-candidates in the FGC in the F850LP-passband. {In other words, more than 80 percent of the original samples by \citet{Jordan2009}, and \citet{Jordan2015}, respectively, are kept in each of our four samples, and the the ones which are left out are mostly near the resolution limit of the HST.}

Note that \citet{Jordan2009} and \citet{Jordan2015} requested that an object must have $r_{\rm h}<10 \, pc$ averaged over the F475W-passband and the F850LP-passband, in order to become a member in their catalogues for GC-candidates in the VGC, and the FGC, respectively. There is however no limit in magnitudes. This means that the objects could well fulfill the criterion according to which a UCD should have a mass of $2 \times 10^6 \, {\rm M}_{\odot}$ or above \citep{Mieske2008}. If a mass-to-light ratio of $1.45 \ {\rm M}_{\odot}/L_{\rm F850LP}$ is set \citep{Sivakoff2007}, then this would correspond to a luminosity $> 10^6 \, {\rm M}_{\odot}$. Indeed, if the distance to Virgo is 16.5 Mpc \citep{Mei2007}, then 485 of the GC-candidates, or about 4.5 percent, have $L_{\rm F850LP} > 10^6 {\rm L}_{{\rm F850LP},\odot}$, and two GC-candidates even have $L_{\rm F850LP} > 10^7 {\rm L}_{{\rm F850LP},\odot}$. And if the distance to Fornax is 20 Mpc (\citep{Blakeslee2009}), then 302 of the GC-candidates, or about 5.6 percent, have $L_{\rm F850LP} > 10^6 {\rm L}_{F850LP,\odot}$ and four even have $L_{\rm F850LP} > 10^7 {\rm L}_{{\rm F850LP},\odot}$. However, many UCDs are more extended than 10 pc, and this becomes the more noticeable the higher the mass of UCD is (see for example  \citealt{Dabringhausen2012}). Thus, only the more compact UCDs can be included in the two optical catalogues.

Also note that the optical data is as uniform as possible. This is done by taking data that was retrieved with the same instrument, and processed the same way into catalogues. Indeed, what \citet{Jordan2015} do for the FGC is basically just a repetition of what \citet{Jordan2009} did for the VGC.

\subsection{X-ray data}
\label{sec:x-ray}

We used the X-ray sources from the second edition of the Chandra X-ray Observatory (CXO) source catalogue \citep{Evans2010}. An overview of the telescope itself is given in \citet{Weisskopf2000}.

The CXO source catalogue (CSC) is not an all-sky catalogue, meaning that there also areas on the sky which are not included in it. It also implies that the observations are not equally deep for each and every source, but depend on how long the CXO was pointed in a given direction. However, those problems also existed if we tried ourselves to combine the single pointings of the CXO into a catalogue of more or less comparable X-ray sources, as for instance \citet{Sivakoff2007} did before the advent of the CSC. The data in the CSC are thus more homogeneous than the raw data, at the potential expense that some data may be missing in comparison to single pointings of the CXO. We will return to this in Section~\ref{sec:sivakoff2007}.

The key parameter here is the resolution of the Chandra X-ray observatory (CXO), because the accreting NSs and BHs are nearly point sources at the distance of the FGC and the VGC. For the CXO, at least half of the flux goes through a circle of one arcsec in diameter for such a source. This is however precise enough to tie most X-ray point sources to a single GC at the distance of the VGC and the FGC.

We selected the X-ray sources that are within two degrees of the central galaxy M87 for the VGC and the X-ray sources that are within two degrees of the central galaxy NGC1399 for the FGC. Thus, at the time of the data retrieval in October 2019, we were taking the widest fields that we could in single pointings for both galaxy clusters. We thereby captured for both clusters the central parts of the optical catalogues, which contain 35 galaxies for the VGC and 12 galaxies for the FGC. 

We thereby found 2639 X-ray sources in the VGC and 1573 X-ray sources in the FGC. Most of them are indeed point-like within the resolution limits of the Chandra X-ray observatory. Those which are not, probably get sorted out in the next step, which is the search of a GC-counterpart in the optical, if there is any. Extended X-ray sources usually do not match with the equally point-like GCs at the distance of the VGC and the FGC, if seen from an X-ray perspective.

\subsection{Optical and X-ray data combined}
\label{sec:combined}

\begin{table}
	\caption{\label{tab:virgo-final} The number of the galaxy in the Virgo Galaxy Cluster Catalogue by \citet{Binggeli1985}, its Johnson-Cousins B-band magnitude, the corresponding B-band luminosities for a distance of 16.5 Mpc \citep{Mei2007} in logarithms to the base of ten of Solar units, the observed GCs around the galaxy in the ACSVCS and the number of them which have an X-ray source. The data on the B-band magnitudes were taken from the HyperLeda database \citep{Makarov2014} found at the webpage http://leda.univ-lyon1.fr.}
	\centering
	\vspace{2mm}
	\begin{tabular}{rrrrr}
		\hline
		&&&& \\ [-10pt]
		VCC...	& ${\rm m}_B$   & $L_B$  & \# of GCs & with X-rays \\
		\hline
		1316 & 9.53  & 10.81 & 1745 &  92 \\
		881  & 9.61  & 10.78 & 367  &  8  \\
		763  & 9.82  & 10.70 & 506  &  21 \\
		1632 & 10.50 & 10.43 & 456  &  42 \\
		1231 & 10.89 & 10.27 & 254  &  8  \\
		1154 & 11.09 & 10.19 & 192  &  4  \\
		1030 & 11.39 & 10.07 & 176  &  2  \\
		1664 & 11.78 & 9.91  & 146  &  5  \\
		759  & 11.85 & 9.89  & 172  &  4  \\
		1242 & 12.24 & 9.73  & 116  &  1  \\
		1619 & 12.27 & 9.72  & 70   &  2  \\
		1327 & 12.62 & 9.58  & 173  &  3  \\
		1125 & 12.65 & 9.57  & 62   &  0  \\
		1630 & 12.66 & 9.56  & 53   &  3  \\
		1537 & 12.67 & 9.56  & 45   &  1  \\
		828  & 12.77 & 9.52  & 80   &  0  \\
		1146 & 12.79 & 9.51  & 82   &  1  \\ 
		1250 & 12.95 & 9.45  & 54   &  0  \\
		1283 & 13.35 & 9.29  & 66   &  0  \\
		1261 & 13.54 & 9.21  & 46   &  0  \\
		1087 & 13.82 & 9.10  & 68   &  2  \\
		1297 & 14.12 & 8.98  & 152  &  1  \\
		1695 & 14.30 & 8.91  & 22   &  0  \\
		1431 & 14.31 & 8.90  & 71   &  0  \\
		1528 & 14.54 & 8.81  & 49   &  0  \\
		1355 & 14.92 & 8.66  & 20   &  0  \\
		1545 & 14.95 & 8.65  & 63   &  0  \\
		1499 & 15.00 & 8.63  & 35   &  0  \\
		1407 & 15.01 & 8.62  & 60   &  0  \\
		1627 & 15.20 & 8.55  & 8    &  0  \\
		1185 & 15.37 & 8.48  & 33   &  0  \\
		1512 & 15.58 & 8.39  & 15   &  1  \\
		1539 & 15.76 & 8.32  & 43   &  0  \\
		1489 & 15.87 & 8.27  & 22   &  0  \\
		&&&& \\ [-10pt]
		\hline
	\end{tabular}
\end{table}

\begin{table}
	\caption{\label{tab:fornax-final} The number of the galaxy in the Fornax Galaxy Cluster Catalogue by \citet{Ferguson1989}, its Johnson-Cousins B-band magnitude, the corresponding B-band luminosities for a distance of 16.5 Mpc \citep{Blakeslee2009} in logarithms to the base of ten of Solar units, the observed GCs around the galaxy in the ACSFCS and the number of them which have an X-ray source. The data on the B-band magnitudes were taken from the HyperLeda database \citep{Makarov2014}.}
	\centering
	\vspace{2mm}
	\begin{tabular}{rrrrr}
		\hline
		&&&& \\ [-10pt]
		FCC...	& ${\rm m}_B$   & $L_B$ & \# of GCs & with X-rays \\
		\hline
		213 & 10.35 & 10.65 & 1075 &  68 \\
		167 & 10.79 & 10.48 & 424  &  15 \\
		219 & 10.81 & 10.47 & 380  &  18 \\
		184 & 11.69 & 10.12 & 306  &  9  \\
		276 & 11.70 & 10.11 & 362  &  32 \\
		147 & 11.93 & 10.02 & 320  &  0  \\
		170 & 12.34 & 9.86  & 71   &  1  \\
		193 & 12:36 & 9.85  & 48   &  1  \\
		153 & 12.90 & 9.63  & 53   &  0  \\
		148 & 13.10 & 9.55  & 87   &  0  \\
		177 & 13.20 & 9.51  & 70   &  0  \\
		190 & 13.71 & 9.31  & 156  &  1  \\
		255 & 13.71 & 9.31  & 80   &  0  \\
		277 & 13.77 & 9.29  & 42   &  0  \\
		301 & 13.96 & 9.21  & 21   &  0  \\
		143 & 14.03 & 9.18  & 63   &  0  \\
		136 & 14.43 & 9.02  & 25   &  0  \\
		95  & 14.48 & 9.00  & 21   &  0  \\
		182 & 14.73 & 8.90  & 59   &  1  \\
		202 & 14.74 & 8.90  & 231  &  12 \\
		100 & 15.21 & 8.71  & 34   &  0  \\
		203 & 15.38 & 8.64  & 31   &  0  \\
		288 & 15.40 & 8.63  & 18   &  0  \\
		106 & 15.56 & 8.57  & 15   &  0	 \\
		&&&& \\ [-10pt]
		\hline
	\end{tabular}
\end{table}

For the VGC, we find that 34 of the 100 galaxies in the ACSVCS are within two degrees of M87, besides M87 itself. All of them are also covered in the Chandra X-ray source Catalogue (CSC). For the FGC, there are a bit more than 20 of the 43 galaxies in the ACSFCS within two degrees of NGC1399. However, only 11 of them, and NGC1399 itself, are also covered in the CSC. This may be an example for that the CSC does not cover the complete sky (see Section~\ref{sec:x-ray}).

The next step is to identify those GC-candidates in the 35 galaxies in the VGC, and 12 galaxies in the FGC, respectively, which are matched by an X-ray source. This is done with \textsc{Topcat}, a computer programme for operations on catalogues and tables by Mark Taylor \citep{Taylor2005}. In more detail, the procedure is as follows.

The X-ray data come from a single catalogue and we thus use them to set the coordinate system. Thus, they do not have errors in their positions in this setting. The optical data in contrast do have errors, namely a random error that is different from GC to GC and one systematic error that is only different from galaxy to galaxy. The random error comes from measurement uncertainties and rounding errors, and the systematic error comes from the Hubble Space Telescope that was newly oriented for each of the galaxies.

For each of the 35, and 12 galaxies, respectively, a preliminary match was made with \textsc{Topcat}. The requirement for a preliminary match is that the distance between an optical source and an X-ray source is less than 0.0006 degrees, or 2.16 arcsec. Then the matches were inspected by eye, in order to find a systematic offset. In most cases there were none detected, either because there really are none, or the matches were too few to find systematics. If an offset was found, it was corrected by changing the coordinates of the optical sources by the appropriate amount with \textsc{Topcat}. Then the final matches were made, which require that a GC-candidate and `its' X-ray source are closer together than 0.0003 degrees, or 1.08 arcsec. These objects are then interpreted as GCs with a bright LMXB.

This two-step method of finding matches between optical and X-ray sources may seem a bit awkward, but it is not useless. For instance, 77 pairs were found for NGC1399 in the first go, but only 69 pairs appear in the final list of GCs with an LMXB. Thus, the final list was cleansed from 8 pairs, or 10 percent of the pairs, that we interpreted as accidental. If, for instance, our method was just to find all pairs of optical and X-ray sources that are closer together than 2.16 arcsec, then we would also have about ten percent false matches. If, on the other hand, we would just look for all pairs of optical and X-ray sources that are closer together than 1.08 arcsec from the beginning, then the real pairs of optical and X-ray sources in NGC1399 with its initial offset of 1.8 arcsec would be rather the exception than the rule.

We are aware of the fact that by applying our two-step method, we might miss some real matches between optical and X-ray data. This would happen in galaxies where also the preliminary match yields $\apprle 5$ pairs or less, so that it remains inconclusive in which direction the optical sources would have to be corrected. Thus, they would not be corrected and thereby miss the final criterion for a match, namely being closer together than 1.08 arcsec. However, we are confident that this happens rarely, because usually matches within 2.16 arcsec are also matches within 1.08 arcsec for the galaxies poor in matches. Or in other words, the 1.8 arcsec by which the optical sources of NGC1399 have to be corrected are probably exceptional in the sample of galaxies used here.

We also thereby note that the random errors of optical and X-ray sources, that is the errors that change between sources instead of between galaxies, are much better than 1.08 arcsec. In the other extreme case, namely that all pairs between optical and X-ray sources are purely accidental, the final matches would be about 25 percent of the initial matches, and not about 90 percent. This is because also the area covered by the criterion for a final match is 25 percent of the criterion for an initial match.

We nevertheless make a crosscheck for the statement that the X-ray sources really are part of the GCs. For this reason, consider five fake corrections of the catalogues with the optical sources. For the first of these fake corrections, 43.2 arcsec was subtracted from the actual declinations in the optical catalogues. For the second to the fifth of the fake corrections, 21.6 arcsec was added each time to the declinations of the GC-candidates. Note that for the third fake correction, there is no change to the declination compared to the raw data. Also note that the ACS captures squared fields with approximately 200 arcsec in length. Thus, the first and the fifth fake correction are about one quarter of the length of an ACS field. For a match between optical data and X-ray data on the other hand, only a distance of less than 2.16 arcsec is required, that is one tenth of the fake corrections.

We let \textsc{Topcat} search for matches between the X-ray catalogue and the optical catalogues with the fake corrections. For the FGC, \textsc{Topcat} finds 69, 78, 192, 74 and 65 matches for fake corrections in declination of -43.2, -21.6, 0.0, 21.6 and 43.2 arcsec. For the VGC, \textsc{Topcat} finds 128, 146, 263, 141 and 128 matches for fake corrections in declination by -43.2, -21.6, 0.0, 21.6 and 43.2 arcsec. From these values, slopes of how many LMXBs are expected per distance can be obtained. For this, first averages between the positive and negative fake corrections for the declination are made; that is for example the matches for the fake declinations of -43.2 and 43.2 arcsec for the VGC are summed, and then divided by two. Then it is evaluated by how much the two averages differ, and for the error, the square root of the average is taken. From this, it is derived that in the VGC the number of LMXBs decreases by $14.5\pm 3.7$ per step size away from zero in declination, and in the FGC it is $9\pm 3$. Thus, the expected number of LMXBs for the uncorrected declination is $159 \pm 4$ for the VGC, and $85 \pm 3$ for the FGC. The real numbers of LMXBs for uncorrected declinations are however significantly higher, namely by 104 for the VGC and by 121 for the FGC.

This number of probable random matches is surprisingly high. This may be due to the admittedly somewhat coarse method, because the preliminary condition for a match is taken. This is because the match here is made with the raw data. This is simpler, because the raw data are just in a single file for all galaxies, while for the final data, there is a file for each galaxy, because they are corrected individually. However, let us suppose the final  conditions for a match were taken. Then the numbers of random matches would probably be lower by a factor of four, that is about 40 for the VGC and 20 for the FGC. Thus, the fraction of random matches would be 40/(104+40)=0.28 for the VGC and 20/(121+20)=0.14 for the FGC. The real numbers are probably better still, because the corrections to the optical data are not included here.

For the final matches, there are 12763 GC-candidates in the VGC, of which 202 coincide with an X-ray source. For the FGC, there are 6275 GC-candidates, of which 157 coincide with an X-ray source. For the samples without the most uncertain optical data (see Section~\ref{sec:optical}), there are for the VGC 10806 GC-candidates in the F475W-passband, of which 181 coincide with an X-ray source, and 10774 GC-candidates in the F850LP-passband, of which 180 coincide with an X-ray source. For the FGC, there are 5397 GC-candidates in the F475W-passband, of which 129 coincide with an X-ray source, and 5354 GC-candidates in the F475W-passband, of which 132 coincide with an X-ray source. 

These X-ray sources are distributed over 18 galaxies in Virgo and 10 galaxies in Fornax. Thus, especially in Virgo, there are many galaxies in whose directions there are also CSC-data, but for which no match between optical and X-ray data is found in the end. Also for the galaxies that do have GC-candidates with an X-ray source, their numbers are most uneven. About half of the total number of LMXBs in either of the two galaxy clusters is concentrated in the most massive central galaxy (M87, and NGC1399, respectively), but there are also galaxies that harbour just one or a few LMXBs. Further information on the galaxies considered, how many GCs are observed around them and how many X-ray sources are found in them is summarized for the VGC in table~\ref{tab:virgo-final} and for the FGC in table~\ref{tab:fornax-final}.

\subsection{Comparison with Sivakoff et al. (2007)}
\label{sec:sivakoff2007}

In a final cross-check, the results from Section~\ref{sec:combined} for the GC-candidates with an X-ray source is compared with the same data from \citet{Sivakoff2007}, which assembled the data from single pointings of the CXO, and carefully fused them also into a catalogue, see specifically their section~2.2. This is to exclude that deep data are lost on the way to make them part of the CSC.

First, the full optical sample of 12763 GC-candidates in the VGC from \citet{Jordan2009} has to be cut to the GC-candidates within two degrees of the centre of M87, that is, the region where the X-ray data is actually considered. 35 galaxies with a total of 5577 GC-candidates fulfill this criterion. Note that the restrictions from Section~\ref{sec:optical} regarding the radius and its error and the error of the magnitudes are not made here, in contrast to the reminder of this paper. Thus, also the GC-candidates with an X-ray source have to be considered unrestricted, which results into 202 objects. In other words, about 3.6 percent of the GC-canditates within two degrees of M87 contain an X-ray source in the CSC.

\citet{Sivakoff2007} discovered 270 X-ray sources in a total of 6658 GC-candidates, which was the largest sample GC-candidates with an X-ray source back then. Thus, 4.1 percent of the GC-candidates contain an X-ray source in \citet{Sivakoff2007}, which is, as of the year 2022, comparable to the CSC in relative and absolute numbers. That the relative number in \citet{Sivakoff2007} is still a bit higher than in our search may be because \citet{Sivakoff2007} took only the 10 brighest galaxies in the VGC and one extra galaxy of comparable size into account, while we have no such limitations on the brightness of the galaxies. It is however known that the GCs of brighter galaxies also tend to be redder on average, as was discovered by \citet{Forbes1997} for the red sub-population and by \citet{Strader2004} for the blue sub-population; see also Section~\ref{sec:GCs}. The LMXBs in turn are more frequent in redder GCs (for instance \citealt{Fabbiano2006}; see also Section~\ref{sec:LMXBtheory}).

In the FGC, there are 157 X-ray sources in the CSC which cooincide with the 3992 optical GC-candidates within two degrees of NGC1399 from \citet{Jordan2015}. This means that about 3.9 percent of the total number of GC-candidates have an X-ray source. Thus, both the absolute and the relative number of GC-canditates with an X-ray source in the FGC are comparable with such numbers for the VGC in \citet{Sivakoff2007} and in this paper. The data from the CSC can therefore be as complete as the data from individual studies, but the advantage is that they are likely to be more homogeneous.

\section{The probability to find an LMXB in a GC}
\label{sec:probability}

\subsection{The theory}
\label{sec:LMXBtheory}

LMXBs are interesting for our study, because they are usually not born as such, but evolve into LMXBs through stellar dynamical processes. Thus, they occur often in star clusters, but almost never in the field, because stellar dynamical processes are an exception in the field. A stellar dynamical process that could lead to the creation of an LMXB are tidal captures of stars \citep{Clark1975,Verbunt1987}.

The frequency of such encounters is measured by the encounter rate, which can be written as
\begin{equation}
\label{eq:encrate1}
\Gamma \propto \frac{n_{\rm s} \, n_{\rm ns} \, r_{\rm c}^3}{\sigma}
\end{equation}
\citep{Verbunt2003}. In this equation, $n_{\rm s}$ is the number density of stars, $n_{\rm ns}$ is the number density of NSs and BHs, $r_{\rm c}$ is the core radius of the particular globular cluster and $\sigma$ the velocity dispersion of its constituents. If the initial stellar mass function is the same in all globular clusters (e.g. \citealt{Kroupa2001}), then $n_{\rm s} \propto n_{\rm ns} \propto \rho_0$, where $\rho_0 \propto M \, r_{\rm c}^{-3}$ is the central density of the GC, and $M$ is its total mass. However, $r_{\rm c}$ is difficult to measure for GCs at the distance of the VGC and the FGC, as they are barely resolved with current instruments. The projected half-light radius $r_{\rm h}$, and thus the half-mass radius under the assumption that mass follows light, is larger and therefore less difficult to retrieve from the data. For practical purposes, it is therefore useful to assume $r_{\rm h} \propto r_{\rm c}$ and $\rho_0 \propto M \, r_{\rm h}^{-3}$. Finally, if the GC is also in virial equilibrium, which it usually is, then $\sigma \propto \rho_0^{0.5}\, r_{\rm h} \propto M^{0.5} \, r_{\rm h}^{-0.5}$. Thus, equation~\ref{eq:encrate1} becomes
\begin{equation}
\label{eq:encrate2}
\Gamma_{\rm h}=\frac{M^{1.5}}{r_{\rm h}^{2.5}}
\end{equation}
with these assumptions. The subscript h indicates that $\Gamma$ is derived from $r_{\rm h}$ instead of $r_{\rm c}$ \citep{Sivakoff2007,Dabringhausen2012}.

The validity of equation~\ref{eq:encrate2} depends of course on whether the assumptions, on which it is based, are fulfilled. \citet{McLaughlin2000} showed for instance that $r_{\rm h}$ is not strictly proportional to $r_{\rm c}$ in GCs in the Milky Way, but that the GCs are on average more concentrated with rising mass. If this is also the case for the galaxies in the VGC and the FGC, then the relation may break down. On the other hand, \citet{Jordan2004} used the more uncertain values for $r_{\rm c}$ for M87, the central and most massive galaxy in the VGC, and \citet{Sivakoff2007} stated that all their conclusions would have remained the same within their errors if \citet{Jordan2004} had used $r_{\rm h}$ instead. We therefore also use estimates on the probability to contain an LMXB based on $r_{\rm h}$ rather than $r_{\rm c}$, also to be able to use the catalogues by \citet{Jordan2009} and \citet{Jordan2015} without further assumptions and modifications.

The probability to find an LMXB in a GC depends also on its colour (see e.g. \citealt{Angelini2001}; for a review section~3.4.3 in \citealt{Fabbiano2006}). Red GCs have about 3 times as many LMXBs as blue GCs \citep{Kundu2002,Jordan2004}. The colour of the GCs is interpreted here as a metallicity effect, while age could have in principle the same effect (see \citep{Worthey1994} for the age-metallicity dependency). Thus, like in \citet{Sivakoff2007}, also the overabundance of LMXBs in red GCs is a metallicity effect here.

\citet{Sivakoff2007} thus assume that the probability to find an LMXB in a GC depend on $\Gamma$ and colour, to degrees which have to be found in a fit. They assume Poisson statistics, which means that the probability to have at least one LMXB in the GC is
\begin{equation}
\label{eq:probability}
P_{X}=1-e^{-\lambda}
\end{equation}
because the probability to have not even one LMXB is $P_{nX}=e^{-\lambda}$. \citet{Sivakoff2007} maximized the logarithm of the likelihood for the GCs of the 10 brightest galaxies from the ACSVCS, and one galaxy similar to these. They find
\begin{equation}
\label{eq:gamma}
\lambda = A \, \Gamma_h^{\alpha} \, Z^{\beta},
\end{equation}
with $\alpha=0.82$ if $\Gamma_{\rm h}$ is divided by $10^7$, $\beta=0.39$ if $Z$ is given in Solar metallicities and $A=0.041$. But of course $A$, $\alpha$ and $\beta$ may be different with different optical and X-ray data.

Thus, the parameters of the GCs that we are interested in, namely their masses, their radii and their metallicities, depend on a single and relatively easy-to-measure quantity, namely whether they have a bright X-ray source or not. Generally, the probability to find an LMXB in a GC rises the more massive the GC is, the smaller its radius is, and the higher its metallicity is. However, it may also play a role how old the GC-system is, how the distribution of its radii are, and so on. Also, the GC-systems could have been observed differently, although we did in this paper our best to exclude this possibility (see Section~\ref{sec:data}). Therefore, it would not be unexpected if the parameters obtained for the VGC do not fit for the FGC. On the other hand, if they do match, it means that their differences do not matter for their GC populations, if they are not identical.

The modelling of the LMXB-frequencies in GC-candidates in the VGC and the FGC in dependency of their masses, radii and metallicities, and the comparison to the real data was done on a computer, as described in the next subsection.

\subsection{The programme}
\label{sec:LMXBprogramme}

The computer programmes that decide whether a GC should theoretically have a LMXB or not work as follows.

First, an optical sample like the ones in Section \ref{sec:optical} is read completely. Thus, information of the luminosities in the F475W-passband and F850LP-passband and the radii are obtained for all GCs in the sample. Note that there are two radii available for each GC, namely one for each passband that was observed. However, here the same passband is chosen for the radii that is also chosen for the estimate for the luminosity and the mass of the GC.

To calculate $\Gamma_{\rm h}$, the physical radii are needed instead of the apparent ones. Thus the apparent radii have to be converted into physical ones with distance estimates. For this step, all distances to the GCs in the VGC are set to 16.5 Mpc \citep{Mei2007} and all distances to the GCs in the FGC are set to 20.0 Mpc \citep{Blakeslee2009}. Thus, we choose the same distances as \citet{Jordan2009}, and \citet{Jordan2015}, respectively.

Also the colours of the GCs are calculated. They are needed to estimate the metallicity from them as
\begin{equation}
\label{metallicity-a}
[Z/{\rm H}]=-2.75+1.83\times({\rm F475W-F850LP})
\end{equation}
if (F475W-F850LP)$<1.05$ and 
\begin{equation}
\label{metallicity-b}
[Z/{\rm H}]=-6.21+5.14\times({\rm F475W-F850LP})
\end{equation}
if (F475W-F850LP)$\ge 1.05$ for the two passbands in the same GC \citep{Peng2006}. The values for $[Z/{\rm H}]$ are transformed into values for $Z$ with $Z=0.02\times 10^{[Z/{\rm H}]}$. There is a cutoff when $Z>0.1$, that is about five times the solar metallicity, while GCs are usually far below Solar metallicity. The program sets then $Z=0.1$. This is to avoid exceptional GCs to spoil the general picture, and it is doubtful whether those very rare exceptions are not just errors in the measurements.

The masses of the GCs are calculated as
\begin{equation}
\label{eq:lum-to-mass}
M=1.45 \, L
\end{equation}
where $M$ is the mass of the GC in Solar units and $L$ is the luminosity in Solar units, independent whether it is measured in the F475W-passband or in the F850LP-passband. For the motivation of using equation~\ref{eq:lum-to-mass} independent of the passband, see \citet{Sivakoff2007}.

Then $\Gamma_{\rm h}$ is calculated for each GC with equation~\ref{eq:encrate2}. This number is, as a first rough approximation to the final value, divided by $10^7$ (see also Section~\ref{sec:LMXBtheory}). Together with the metallicity, $\gamma$ is calculated with $\alpha=0.82$, $\beta=0.39$ and $A=1$ (cf. equation~\ref{eq:gamma}). Based on this, the first guess of the theoretical probability, $P_{\rm theo}$, is made with equation~\ref{eq:probability}. When summing up $P_{\rm theo}$ for all GCs in the optical sample, the result should be the total number of GCs with at least one LMXB in the sample, $\sum P_{\rm theo}$, but generally it is not. This issue is solved by rescaling $\lambda$ in equations~\ref{eq:probability} and~\ref{eq:gamma} in each GC by the same amount, until $\sum P_{\rm theo} \approx N_{\rm LMXB}$ is reached. $N_{\rm LMXB}$ is the actual total number of LMXBs in the sample of GCs. In practice, this is done with the Newton-Raphson method for finding a root to the function $f=\sum P_{\rm theo}-N_{\rm LMXB}$. The iterations of this method are stopped if $f\times f \leq 10^{-3}$, and the resulting value for $A$ is taken.

Then a theoretical population of GCs with at least one LMXB is created, based on the optical data for the GCs and the parameters $\alpha$, $\beta$ and $A$. For this, first a random number between 0 and the total number of GCs in the optical sample is diced. If the number is between $n-1$ and $n$, the $n$th GC is chosen. The value for $\lambda$ of that GC is calculated with equation~\ref{eq:gamma}, and then the probability of the GC to have at least one LMXB is calculated with equation~\ref{eq:probability}. Then a second random number is diced between 0 and 1. If this number is below the probability just calculated, the luminosity of the GC is written into a data file. Note that the mass of the GC is actually needed to calculate the probability for it to have an LMXB, but it is linked to its luminosity with equation~\ref{eq:lum-to-mass}. The programme also raises the count of GCs with an LMXB by one. If the random number is larger than the probability, the GC is discarded without any further actions. This process is repeated until $1.6\times 10^6$ GCs with an LMXB are found, independent of how many GCs in total have been looked at to reach this number. After $1.6\times 10^6$ GCs with an LMXB have been created, they are sorted by their luminosities.

Note that only luminosities that appear in the optical sample can appear in the list of GCs that have an LMXB. However, how often a GC with a certain luminosity shows up in that list depends on how probable it is that it has an LMXB. It is not dependent on whether it has an LMXB in reality.

Cumulative probabilities are built from of the theoretical list of luminosities of the GCs that have an LMXB, as well as the list of GCs that have an LMXB in reality. The cumulative probabilities are functions that have the luminosity of the GCs on the x-axis, raise by a fixed amount for each GC with at least one LMXB on the y-axis, and are normalised to one. The closer two cumulative probabilities are together, the more likely it is that they are consistent with being drawn from the same parent distribution. This allows us to compare the distribution of GCs with an LMXB in theory with the correspondent distribution in practice.

The actual test is done with the Kolmogorov-Smirnov (KS) test. The KS-test translates the maximum distance between two cumulative probability distributions into a probability for two distributions to be consistent with being drawn from the same (potentially unknown) parent distribution\footnote{The test does {\it not} check for actual probabilities here. Consider for instance a test where three numbers are drawn, put back into the urn again, and then three number are drawn for a second time. Suppose the three numbers are the same in the two cases, \{1,2,4\} say. It could then be that the number were indeed drawn from the same distribution in the urn (i. e. the same parent distribution), but it could also be that the parent distribution was different at each time, \{1,2,3,4,5\} and \{1,2,4,8\}, say. The KS-test delivers in both cases 100 percent, but because the parent distribution could in principle change between the first and the second drawing, only for a {\it consistency} of being drawn from the same parent distribution. If on the other hand \{1,2,4\} were drawn at the first time and \{1,2,8\} at the second time, the consistency with being drawn from the same parent distribution was less than 100 percent. This is because the two trials lead to different results, but because size and content of the parent distribution(s) are in principle unknown, it cannot be excluded that the 8 was among the numbers that were left in the urn in the first trial.}. The probability that the distance between two cumulative distributions is larger than the distance $D$ can be approximated by
\begin{equation}
\label{eq:KStest-1}
Q_{KS}(z)=2\sum_{j=1}^{\infty}(-1)^{j-1} \exp (-2 \, j^2 \, z^2) 
\end{equation}
or
\begin{equation}
\label{eq:KStest-2}
Q_{KS}(z)=1-\frac{\sqrt{2 \, \pi}}{z}\sum_{j=1}^{\infty} \exp \left[ -\frac{(2 \, j -1)^2 \, \pi^2}{8\, z^2} \right]
\end{equation}
for $z>0$, where $z$ is given as
\begin{equation}
\label{eq:KSnumber-1}
z=\sqrt{N}+0.12+\frac{0.11}{\sqrt{N}} \, D.
\end{equation}
$N$ in turn is
\begin{equation}
\label{eq:KSnumber-2}
N=\frac{N_1 \, N_2}{N_1 + N_2},
\end{equation}
where $N_1$ is the number of elements in first cumulative distribution and $N_2$ is the number of elements in second cumulative distribution (e.g. \citep{Press1992}). It follows as a special case from equation~\ref{eq:KSnumber-2} that for $N_1 \ll N_2$, $N_2 \approx N$.

The series in equations~\ref{eq:KStest-1} and~\ref{eq:KStest-2} both converge for $z>0$, but at different speeds. For $z\apprge 1.18$, equation~\ref{eq:KStest-1} is faster, and else equation~\ref{eq:KStest-2} (cf. \citealt{Press1992}). Hence, in our programme, we decide for the faster method accordingly. The result can be interpreted in the following way: If $N_1$ and $N_2$ are indeed drawn from the same parent distribution, then there is a chance of $X$ percent that the maximum distance $d$ is greater than the observed distance, $D$.

We evaluate the series in equations~\ref{eq:KStest-1} and~\ref{eq:KStest-2} only until the 4th summand. However, for the GCs with an LMXB, we compare this method with the method of directly searching for distances greater than $D$ for $10^5$ times. Note that in the case here $N_1 \gg N_2$. Thus, $N_1$ is set to the parent function, and furthermore $N_2 \approx N$. As a result, the percentage of cases with $d > D$ is between 0 and $\approx 25$ percent larger in the case that it is estimated with equations~(\ref{eq:KSnumber-1}) to~(\ref{eq:KSnumber-2}) to the 4th order than in the case that it estimated directly from the $10^5$ results of comparing $d$ with $D$. We choose the comparison with equations~(\ref{eq:KStest-1}) to~(\ref{eq:KSnumber-2}) to the 4th order, because it is 1.) more conservative and 2.) more versatile, as it does not need a known parent function.

\section{Results}
\label{sec:results}

\subsection{Direct comparison of the GCs in the FGC and the VGC}
\label{sec:direct}

\subsubsection{General remarks}
\label{sec:GenRemarks}

We first compare only the optical data of the GCs in the VGC with the GCs in the FGC. In a next step, we combine the optical data with X-ray data. Thus, the probabilities of the GCs to have an LMXB are discussed in this step.

A direct comparison of the optical properties of the GCs in the VGC with those in the FGC is tempting, because no match with X-ray data is necessary. This is therefore not a comparison between the 18 galaxies in the VGC and 10 galaxies in the FGC, for which at least one GC with an X-ray source was detected, but rather between the full samples in the optical, with GCs in 100 galaxies in the VGC and 44 galaxies in the FGC. Thus, the statistics is far better in the latter case.

However, a problem that exists for either case is how uneven the sources are distributed over the galaxies. For instance, in the optical passbands, there are 1745 GC-candidates belonging to M87, but only 11 belonging to VCC33. On the other hand, M87 is the largest contributor, but it is responsible for only about 8 percent of the total catalogue of GC-candidates in the VGC. Likewise, there are 93 X-ray sources belonging to M87, but there are also galaxies that have only a few, or even one X-ray source.

In case of a comparison of absolute magnitudes and radii of the GCs, a drawback is that they are directly dependent on the distance, as further discussed in Sections~\ref{sec:directL} and~\ref{sec:directR}. The colours of the GCs, and based on them the estimates for their metallicities, are on the other hand much less distance-dependent (see Section~\ref{sec:directM}). A dependency on distance also exists for the theoretical formation of LMXBs through the encounter rate (Section~\ref{sec:LMXBtheory}), because the probability for the formation of a LMXB depends on luminosity and radius, too.

Also note that because the metallicity is calculated from the colours of GCs, a GC that appears in a sample selected for the F475W-passband and for the F850LP-passband will have exactly the same metallicity in both bands, in contrast to luminosities and radii. We nevertheless consider selections of the GCs in the F475W-passband and in the F850LP-passband also for the metallicities, because the choice of GCs depends on the passband. For instance, let there be a GC which has a large radius error in the F475W-passband, but a small radius error in the F850LP-passband. It may be sorted out because of that in the sample based on the properties in the F475W-passband, but not in the sample based on the properties in the F850LP-passband. Its colour however, and thus its metallicity estimate, is known independent of the errors in the radius.

\subsubsection{Direct comparison between luminosities of the GCs}
\label{sec:directL}

\begin{figure*}
\centering
\includegraphics[scale=0.95]{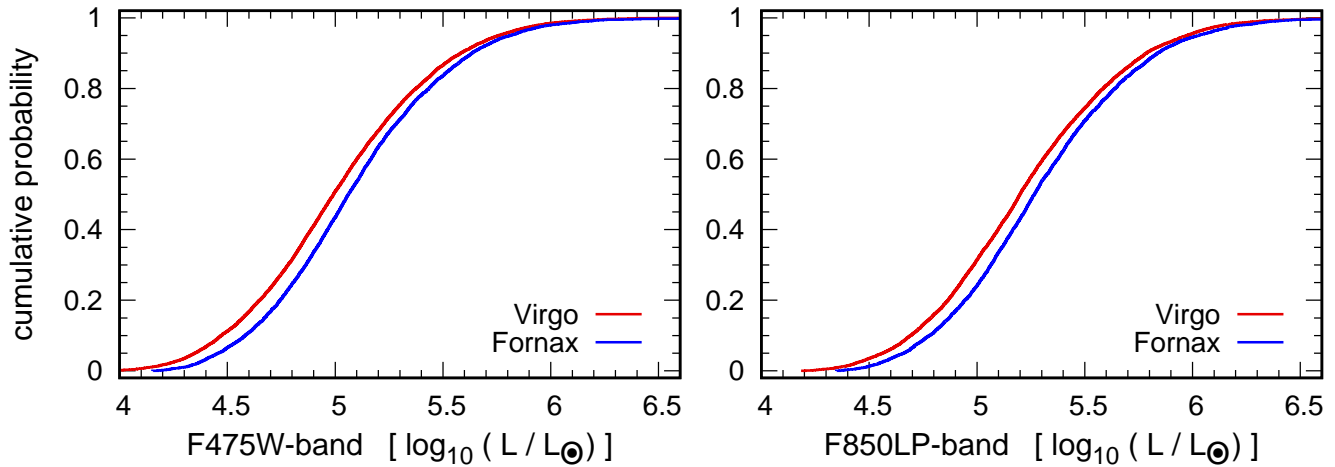}
\caption{\label{fig:comp-lum-full}
Comparisons between the luminosities of the full samples of considered GCs in the VGC (red distributions) and in the FGC (blue distributions). The left panel is for the luminosities in the F475W-passband and the right panel is for the luminosities in the F850LP-passband.}
\end{figure*}

A major problem with the direct comparison of the luminosities and radii of the GCs is that these quantities depend on the distance of the galaxy cluster to which the GC belongs, besides the apparent magnitudes and apparent radii of the GCs themselves. Only if at least the ratio of the distances of the two galaxy clusters is correct (it does not have to be correct in absolute numbers), the luminosities and radii of GCs in VGC can be compared directly with the appropriate values in the FGC.

To illustrate this point, comparisons between the luminosities (Fig.~\ref{fig:comp-lum-full}) and the radii (Fig.~\ref{fig:comp-radii-full}) of the unrestricted samples of the GCs in the VGC and the GC in the FGC are shown.

It can be seen in Fig.~\ref{fig:comp-lum-full} that the luminosities of the GCs in the FGC appear to be somewhat more luminous than in the VGC. This happens in the F475W-passband as well as in the F850LP-passband, and could either indicate differences in the luminosities of the GCs, or be a consequence of the assumption that all GCs in the VGC are 16.5 Mpc away \citep{Mei2007} and all GCs in the FGC are 20.0 Mpc away \citep{Blakeslee2009}. Changing the distance to the from 20.0 Mpc to 19.0 Mpc for the FGC, like in Table~8 in \citet{Freedman2001}, would decrease the estimated luminosities of the GCs in the FGC. The GC systems of the FGC and the VGC would then be closer to have their peaks at the same absolute luminosities, also known as turn-over magnitudes. Note that universal turn-over magnitudes of their GC systems are known as distance estimators for galaxies (e.g. \citealt{Richtler2003}).

Incidentally, also the percentage of GCs that have luminosities $L_{\rm F850LP}>10^6 \, L_{{\rm F850LP}, \odot}$ (and are in the range that makes them to UCD candidates) is at 5.6 percent for a distance of 20.0 Mpc of the FGC, compared to 4.5 percent for the VGC at a distance of 16.5 Mpc. This this percentage would sink to 4.7 percent if the FGC was only 19.0 Mpc away.

Moreover, the distributions in luminosity for the VGC are a bit flatter than the distributions in the FGC. This could be because some bright GCs are actually closer to the observer than the flat distance value adopted here. They thus appear even more luminous, if they are put at a larger distance. In other words, the different steepnesses of the luminosity distributions could be an effect of the 3D-structure of the galaxy clusters. The steeper distributions for the FGC could indicate that the FGC is also more compact along the line of sight than the VGC, and not just projected on the sky.

\subsubsection{Direct comparison between the radii of the GCs}
\label{sec:directR}

\begin{figure*}
\centering
\includegraphics[scale=0.95]{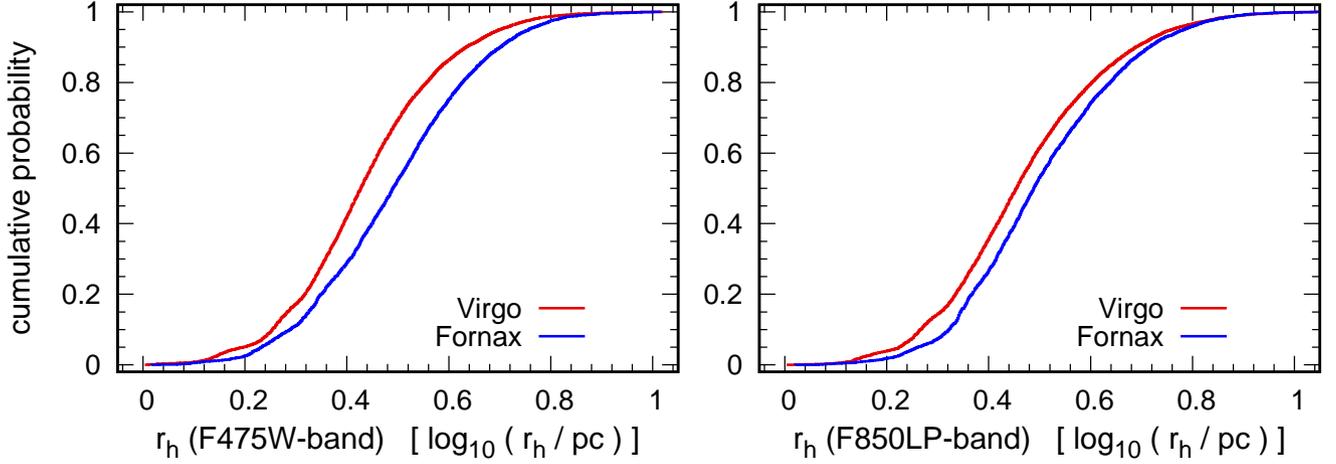}
\caption{\label{fig:comp-radii-full}
Comparisons between the half-light radii of the full samples of considered GCs in the VGC (red distributions) and in the FGC (blue distributions). The left panel is for the half-light radii in the F475W-passband and the right panel is for the half-light radii in the F850LP-passband.}
\end{figure*}

\begin{figure}
\centering
\includegraphics[scale=0.85]{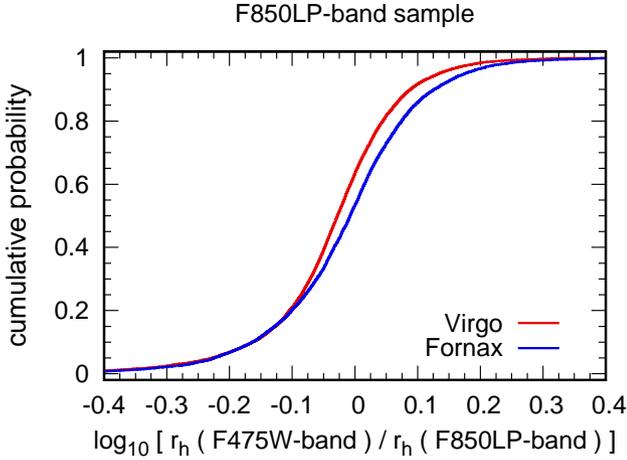}
\caption{\label{fig:comp-radii-ratioz} Comparison of the ratios of the half-light radii of the GC, $r_{\rm h (F475W)} / r_{\rm h (F850LP)}$, in the VGC (red curve) and in the FGC (blue curve) in the sample based on availability of the luminosity in the F850LP-passband. The samples based on the availability of the luminosity in the F475W-passband are almost identical and thus not shown here.}
\end{figure}

Also the radii of the GCs appear to the somewhat larger in both passbands in the FGC than in the VGC, as can be seen in Fig.~\ref{fig:comp-radii-full}. However, the two distributions are more or less parallel in the F850LP-passband. Thus, assuming that both galaxy clusters have the same GC populations, the difference in the F850LP-passband could be resolved by adjusting the distances to the galaxy clusters, so that the two distributions completely cover each other (within errors). This can be done again by replacing the distance of the FGC of 20.0 Mpc with 19.0 Mpc.

However, in the F475W-passband, the distribution for the radii of the GCs in the FGC is flatter than the distribution for the VGC. In other words, the theory that the GC systems from the FGC and the VGC are drawn from the same parent population is far less likely to be true in the F475W-passband than in the F850LP-passband, and the difference of the radii of the GCs in the F475W-passband indicate that they are indeed different populations. Also, the GCs in the FGC are on average still wider than in the VGC in the F475W-passband, even if the smaller value from \citet{Freedman2001} for the distance of the FGC is adopted.

This can be interpreted as that the GCs in the VGC are on average more mass-segregated than the GCs in the FGC, because massive stars emit bluer light. They are thus primarily responsible for the light emission in the F475W-passband. In the F475W-passband however, the GCs in the VGC are denser than the GCs in the FGC, in contrast to the F850LP-passband.

This could indicate that the ratio of the distances between the FGC and the VGC appears to be closer to unity than suggested in \citet{Blakeslee2009}. Thus, if 16.5 Mpc for the distance of the VGC is taken for granted (see \citealt{Mei2007}), then the FGC appears more likely to be 19.0 Mpc away \citep{Freedman2001} than 20.0 Mpc \citep{Blakeslee2009}.

On the other hand, what \citet{Blakeslee2009} actually measure is the distance ratio between the VGC and the FGC. There is however also a distance to the VGC in \citet{Freedman2001}, which they give as 15.3 Mpc. Thus, if the distance measurements to both galaxy clusters are taken from \citet{Freedman2001}, then the distance ratio is almost unchanged. We therefore stick to the distance estimates from \citet{Mei2007} for the VGC and \citet{Blakeslee2009} for the FGC, as in the catalogues used here.

An alternative to measure a possible mass segregation completely independent of distance estimators is to look at the ratios of radii of the GCs in F475W-passband and in the F850LP-passband, as done in Fig.~\ref{fig:comp-radii-ratioz}. This figure confirms that in the VGC, the ratios are on average smaller than in the FGC, which means again that the GCs in the VGC could be more mass-segregated. The consistency of the VGC sample and the FGC sample with being drawn from the same parent distribution is $P<10^{-6}$ according to a KS-test.

\subsubsection{Direct comparison between metallicities of the GCs}
\label{sec:directM}

\begin{table*}
\caption{\label{tab:comp-m} Comparison of the distributions of the metallicities of GCs in the FGC to GCs in the VGC. The first column tells which fraction of the GCs is compared, the second column states in which optical band the GCs are observed, and the third to fifth column state the probability that the distance between the two distributions is even larger than the case realised here according to a KS-test, if both distributions are drawn from the same parent distribution. This means (here and in the following tables) that the higher the displayed number (probability), the more likely the two distributions could have the same parent distribution. Lower numbers rather point to different underlying theoretical distributions. The third column is for all GCs, the fourth only for the GCs with $r_{\rm h} > 2.4\, {\rm pc}$ in the F475W-passband and with $r_{\rm h} > 2.5\, {\rm pc}$ in the F850LP-passband, and the fifth column only for the GCs with $r_{\rm h} \le 2.4\, {\rm pc}$ in the F475W-passband and with $r_{\rm h} \le 2.5\, {\rm pc}$ in the F850LP-passband.}
\centering
\vspace{2mm}
\begin{tabular}{lllll}
\hline
&&&& \\ [-10pt]
GC-sample                     					      		& band   & all radii & large radii & small radii \\
\hline
all GCs                       							& F475W  &  0.00126  &  0.303 	 & 0.0338 \\
all GCs										& F850LP &  0.00277  &  0.108 	 & 0.0274 \\
all GCs, without M87 and NGC1399						& F475W  &  0.000066 &  0.180	 & 0.00371\\ 
all GCs, without M87 and NGC1399                    				& F850LP &  0.000130 &	0.00859  & 0.0243 \\
GCs $ > 5.5 \times \log_{10}(L / {\rm L}_{\odot})$ 	       			& F475W  &  0.694    &  0.638 	 & 0.0344 \\
GCs $\le 5.5 \times \log_{10}(L / {\rm L}_{\odot})$				& F475W  &  0.00121  &  0.325    & 0.0560 \\
GCs $ > 5.5 \times \log_{10}(L / {\rm L}_{\odot})$ 				& F850LP &  0.640    &  0.321 	 & 0.531  \\
GCs $\le 5.5 \times \log_{10}(L / {\rm L}_{\odot})$			      	& F850LP &  0.00245  &  0.0253   & 0.00513\\
GCs $> 5.5 \times \log_{10}(L / {\rm L}_{\odot})$, without M87 and NGC1399	& F475W  &  0.338    &  0.701 	 & 0.470  \\
GCs $\le 5.5 \times \log_{10}(L / {\rm L}_{\odot})$, without M87 and NGC1399	& F475W  &  0.000217 &  0.166    & 0.0132 \\
GCs $> 5.5 \times \log_{10}(L / {\rm L}_{\odot})$, without M87 and NGC1399	& F850LP &  0.0619   &  0.927 	 & 0.113  \\
GCs $\le 5.5 \times \log_{10}(L / {\rm L}_{\odot})$, without M87 and NGC1399	& F850LP &  0.000369 &  0.00364  & 0.0346 \\
&&&& \\ [-10pt]
\hline
\end{tabular}
\end{table*}

\begin{figure}
\centering
\includegraphics[scale=0.85]{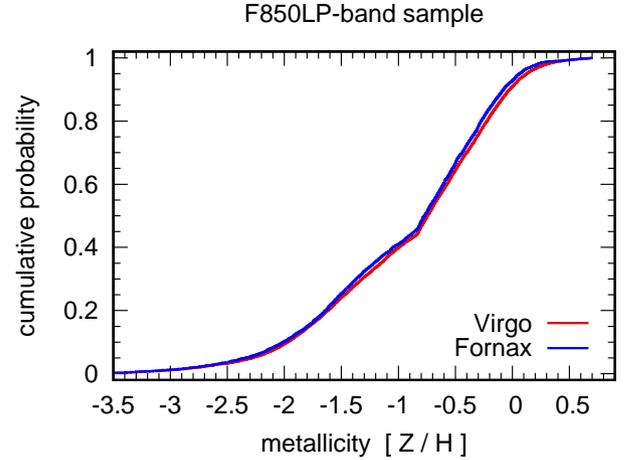}
\caption{\label{fig:virgo+fornax-met-full}
Comparison between the metallicities of the full samples of considered GCs in the VGC (red distribution) and in the FGC (blue distribution). The metallicities are based on the two colours of the GCs, that is if a GC is the sample selected for the F475W-passband and in the sample selected for the F850LP-passband, it will have exactly the same metallicity in both samples. The only difference can be in the selection of the samples, but the samples based on the availability of the luminosity in the F475W-passband are almost identical and thus not shown here.}
\end{figure}

\begin{figure}
\includegraphics[scale=0.85]{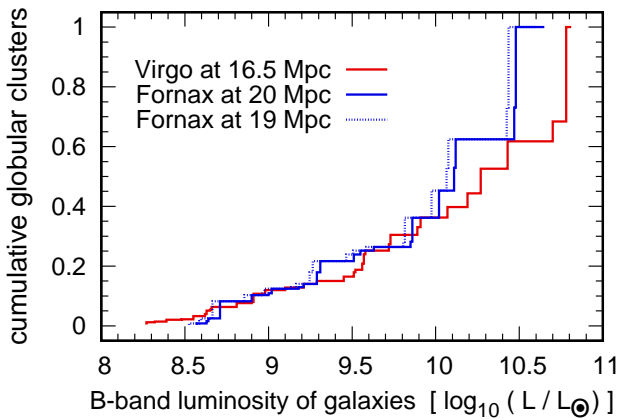}
\caption{\label{fig:lum-number}
The cumulative number of the GCs over the luminosities of their host galaxies for the VGC (red curve) and the FGC (blue curves).}
\end{figure}

The metallicity estimates are not as strongly dependent on distance as the estimates of absolute luminosities or radii. The reason is that they are estimated from the colours, that is in the case here the difference between the apparent luminosities in the F475W-passband and the F850LP-passband for the same GC. A direct distance dependence is therefore ruled out. What possibly remains is a much weaker dependency of colour on distance due to the expansion of the Universe, the so-called redshift. However, the GCs in the FGC are on average a bit bluer than the GCs in the VGC. This is although the FGC is a bit more distant, that is redder than the VGC according to red shift.

Figure~\ref{fig:lum-number} shows that especially the most luminous galaxies in the FGC are less luminous than the most luminous in the VGC. Thus, this is very likely the well-known problem that less luminous galaxies have on average also bluer GC-systems (for instance \citealt{Forbes1997,Strader2006}). Replacing 20 Mpc with 19 Mpc for the distance of the FGC, while keeping the VGC at 16.5 Mpc, has here a negligible effect, in contrast to the luminosities (Section~\ref{sec:directL}) or the radii (Section~\ref{sec:directR}) of the GCs.
	
Because they are bluer, the GCs in the FGC could on average be more metal poor than the ones in the VGC. This is shown in Fig.~\ref{fig:virgo+fornax-met-full}, where the translation from colours to metallicities was done with equations~\ref{metallicity-a} and~\ref{metallicity-b}. However, it is also possible that the GCs in the FGC are on average bluer because they are on average younger than the GCs in the VGC. This is because of a degeneracy of the colours of stellar populations with metallicity or age \citep{Worthey1994}.

We perform KS-tests between all considered GCs in the FGC and all considered GCs in the VGC (Table~\ref{tab:comp-m}). This results to a probability of only 0.1 percent in the F475W-passband and 0.3 percent in the F850LP-passband that the populations have the same properties. The probability rise to 30 percent if only the GCs more extended than $r_{\rm h}=2.4$ pc in the F475W-passband are considered. Likewise, the probability rises 11 percent if only the GCs more extended than $r_{\rm h}=2.5$ pc in the F850LP-passband are considered. (For these choices about compact and extended radii, see Section~\ref{sec:compact}.) Also the probabilities for GCs below extensions of $r_{\rm h}=2.4$ pc in the F475W-passband, and $r_{\rm h}=2.5$ pc in the F850LP-passband, respectively, rise. They do this however to a much smaller extent, namely to about 3 percent in either passband.

Likewise, the probabilities rise to 64 percent in the F475W-passband if only GCs more luminous than $5 \times 10^5 \, {\rm L}_{\odot}$ are considered in a KS-test. If only GCs are considered that are more luminous than $5 \times 10^5 \, {\rm L}_{\odot}$ in the F850LP-passband, the probability that the GCs in the VGC have the same properties as the ones in the FGC is 69 percent. The percentages lower again if only extended or compact GCs above a luminosity of $5 \times 10^5 \, {\rm L}_{\odot}$ are considered, but there still is a significant probability that the populations have the same properties. Also when M87 and NGC1399 are excluded from the comparison of the GCs more luminous than $5 \times 10^5 \, {\rm L}_{\odot}$ in the F475W-passband or the F850LP-passband, it cannot be excluded with high significance that the GCs in the VGC and in the FGC have the same properties. 

Thus, in summary, it is especially the compact GCs with low luminosities that cause the differences in between the VGC and the FGC.

\subsection{Comparisons between GCs that have an LMXB in the VGC and the FGC}
\label{sec:lmxb}

\subsubsection{LMXBs in the full GC-samples}
\label{sec:full-lmxb}

\begin{figure*}
\centering
\includegraphics[scale=0.95]{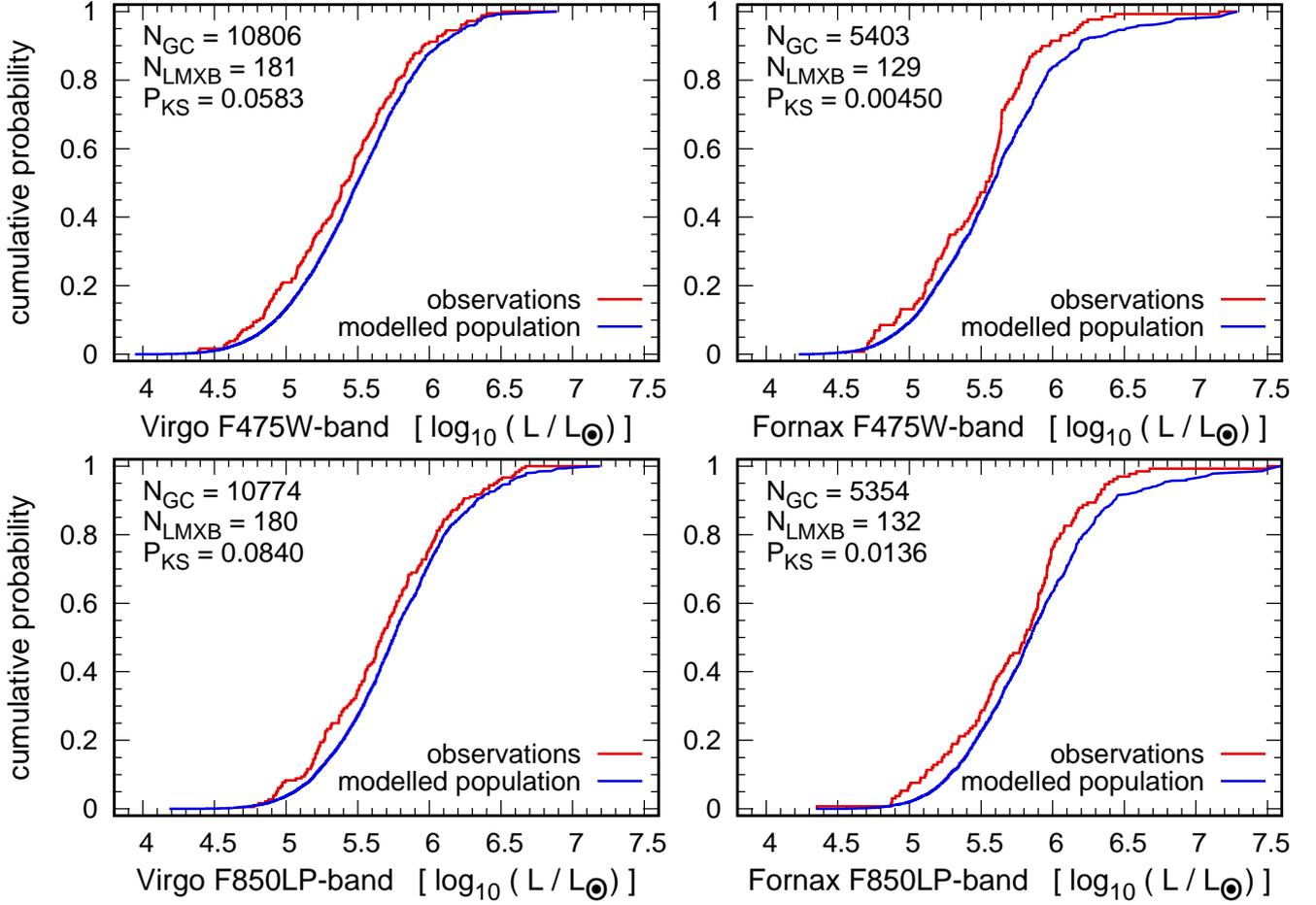}
\caption{\label{fig:full} Comparisons between the actually fully considered samples of GCs with LMXBs and the theoretical predictions. For the theoretical predictions, equation~15 from \citet{Sivakoff2007} was used, or equation~\ref{eq:gamma} here. The nomalisation $A$ however was calculated anew for every comparison, see Section~\ref{sec:LMXBprogramme}. The panels to the left show the comparisons for the VGC and the panels to the right show those for the FGC. Likewise, the comparisons of the top are those for the F475W-passband and the comparisons of the bottom are those for the F850LP-passband. \citet{Sivakoff2007} used their equation~1 for their mass estimates; that is the masses they need to estimate $\Gamma_h$ are based on the F850LP-passband. Noted in each panel are the total number of GCs considered, $N_{\rm GC}$, the number of GCs that have an X-ray source interpreted as LMXB, $N_{\rm LMXB}$, and the probability according to a KS-test that the observed LMXBs are consistent with having formed though encounters involving neutron stars and black holes in the GCs that harbour them, $P_{\rm KS}$ (Section~\ref{sec:probability}).}
\end{figure*}

We perform at first KS-tests for the full samples of GCs in both passbands in the VGC and in the FGC in Fig.~\ref{fig:full}. The samples of GCs that are considered for the search for LMXBs in the VGC contain 10806 GCs in the F475W-passband and 10774 GCs in the F850LP-passband. 180 of them have an LMXB in the F475W-passband and 181 in the F850LP-passband. The correspondent total number of GCs in Fornax is 5403 GCs in the F475W-passband and 5354 GCs in the F850LP-passband. 129 of them have an LMXB in the F475W-passband and 132 in the F850LP-passband.

The probability for Virgo that the distance between the measured distribution and the theoretical distribution is at least as large as observed is close to 6 per cent in the F475W-passband and a bit more than 8 per cent in the F850LP-passband. These fits are moderate at best, but better than the probabilities in Fornax with approximately 0.5 per cent in the F475W-passband and 1.4 per cent in the F850LP-passband.

\subsubsection{Without the GCs in the most massive galaxies}
\label{sec:full-wo-largest-lmxb}

\begin{figure*}
\centering
\includegraphics[scale=0.95]{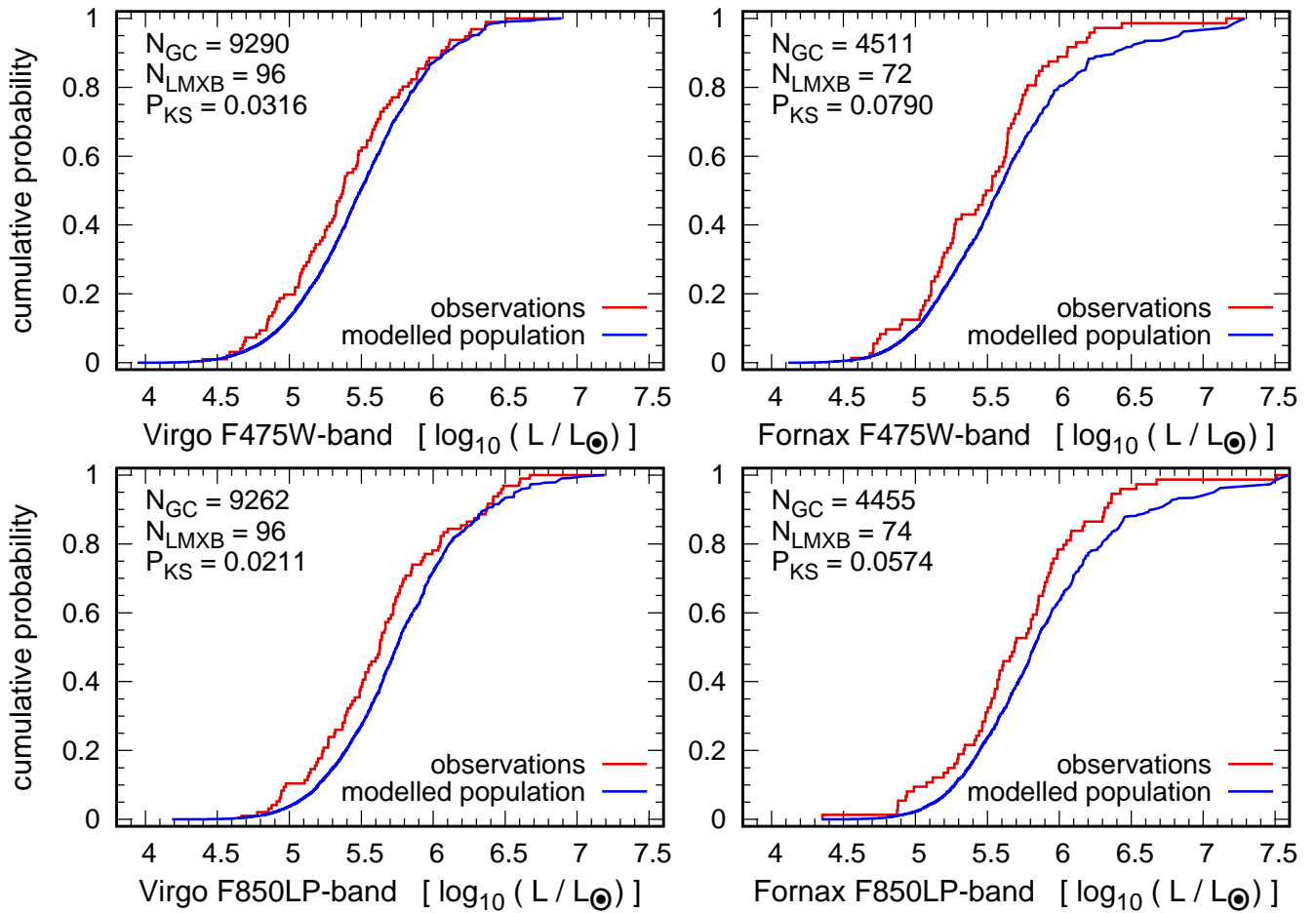}
\caption{\label{fig:full-wo-largest} Like Fig.~\ref{fig:full}, but without the GCs in the most massive galaxies in each galaxy cluster, that is M87 in the VGC and NGC1399 in the FGC.}
\end{figure*}

A bit less than one half of the GCs with an LMXB belong to the galaxies M87 in the VGC, and NGC1399 in the FGC, respectively. These galaxies possess also particularly many compact GCs, where the limit that distinguishes compact and extended GCs is set to 2.4 pc for the F475W-passband and to 2.5 pc for the F850LP-passband in both galaxy clusters (see Section~\ref{sec:compact} for these choices of the radii). Around M87, there are thus 44.3 percent of the GCs compact in the F475W-passband and 39.2 percent of the GCs compact in the F850LP-passband. The percentage of compact GCs in 17 remaining galaxies with at least one GC with an LMXB in the VGC is 34.3 percent in F475W-passband and 34.4 percent in the F850LP-passband. Likewise, there are 37.0 percent of the GCs compact in the F475W-passband and 34.8 percent of the GCs compact in the F850LP-passband, around NGC1399, the most massive galaxy in the FGC. The percentage of compact GCs in 9 remaining galaxies with at least one GC with an LMXB in the FGC is 22.9 percent in F475W-passband and 24.2 percent in the F850LP-passband. Thus, the percentage of compact GCs is higher around the most massive galaxy in either galaxy cluster. This also explains why they contain so many LMXBs in comparatively few GCs, because the encounter rate, $\Gamma$, scales with $r_h^{-2.5}$ (Section~\ref{sec:LMXBtheory}).

To exclude the hypothesis that it is one of those galaxies that is responsible for the differences in the galaxy clusters, we now test GCs in the galaxy clusters without M87, and NGC1399, respectively. The samples of GCs that are considered for the search for LMXBs in the VGC contain now 9290 GCs in the F475W-passband and 9262 GCs in the F850LP-passband. 96 of them in either passband have an LMXB, even though it is not necessarily the same GCs in both passbands. The correspondent total number of GCs in the FGC is 4511 GCs in the F475W-passband and 4455 GCs in the F850LP-passband. 72 of them have an LMXB in the F475W-passband and 74 in the F850LP-passband.

Once again KS-Tests are made in order to compare the theory with the observed LMXBs. However, in order to keep the tests meaningful, the $1.6 \times 10^6$ entries in the theoretical tables have to be calculated anew, before they are compared to the observations. This means in the case at hand that the generating GC-samples for the theoretical expectations are calculated from the catalogues from \citet{Jordan2009} and \citet{Jordan2015} without the data for M87, and NGC1399, respectively. Also in Sections~\ref{sec:5e5-lmxb} and~\ref{sec:5e5-wo-largest-lmxb}, the theoretical samples are each time adjusted to observational sample they are to be compared with. For a more detailed description of how the theoretical samples are generated, see Section~\ref{sec:LMXBprogramme}.

For the VGC without M87 and the FGC without NGC1399, the results are shown in Fig.~\ref{fig:full-wo-largest}. According to the right panels of Fig~\ref{fig:full-wo-largest}, the probability according to a KS-test for the distance being as large or larger as observed if both distributions were drawn from the same parent distribution, $P_{\rm KS}$ raises about one magnitude in both passbands for the FGC compared to the the case when also NGC1399 is included (Fig.~\ref{fig:full}). On the other hand, $P_{\rm KS}$ stays also without NGC1399 well below 10 percent.

For the VGC, $P_{\rm KS}$ is at about 3 percent in the F475W-passband and about 2 percent in the F850LP-passband if M87 is excluded (left panels of Fig.~\ref{fig:full-wo-largest}), and thus even a bit lower than if M87 is included (left panels of Fig.~\ref{fig:full}). However, the change is quite low compared to the VGC with M87. It is also lower than in the FGC with NGC1399 (right panels of Fig.~\ref{fig:full}) compared to the FGC without NGC ((right panels of Fig.~\ref{fig:full-wo-largest})).

Thus, in summary, even though the largest galaxy in either galaxy cluster is special by their number of GCs and by how many of them are compact, leaving them out does little to resolve the issue of the low probabilities, $P_{\rm KS}$, of producing all LMXBs in the remaining galaxies with encounters. 

\subsubsection{Dividing the LMXB-samples at $5.5 \times \log_{10}(L / {\rm L}_{\odot})$}
\label{sec:5e5-lmxb}

\begin{figure*}
\centering
\includegraphics[scale=0.95]{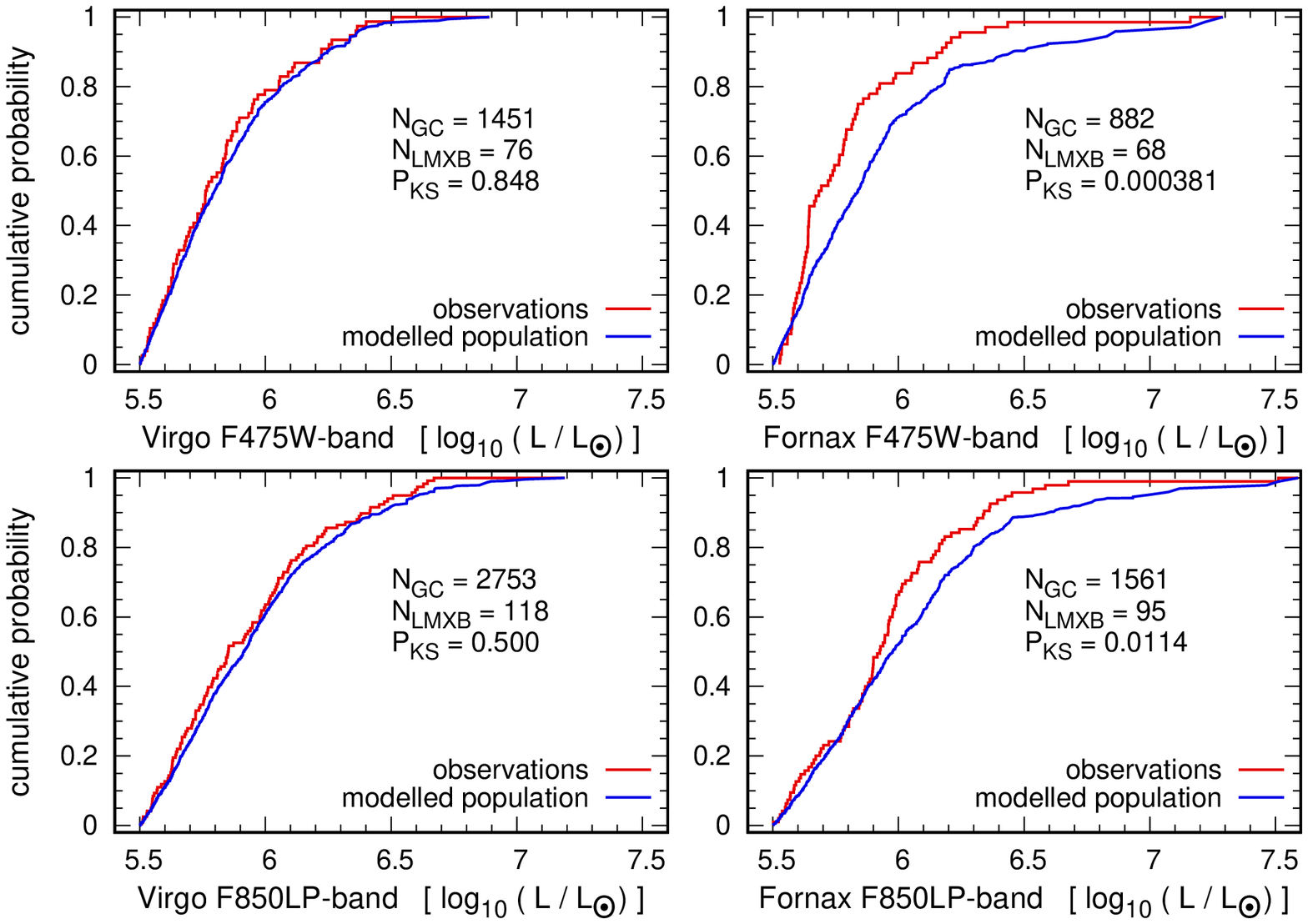}
\caption{\label{fig:g5e5} Like Fig.~\ref{fig:full}, but only for GCs more luminous than $5.5 \times \log_{10}(L / {\rm L}_{\odot})$.}
\end{figure*}

\begin{figure*}
\centering
\includegraphics[scale=0.95]{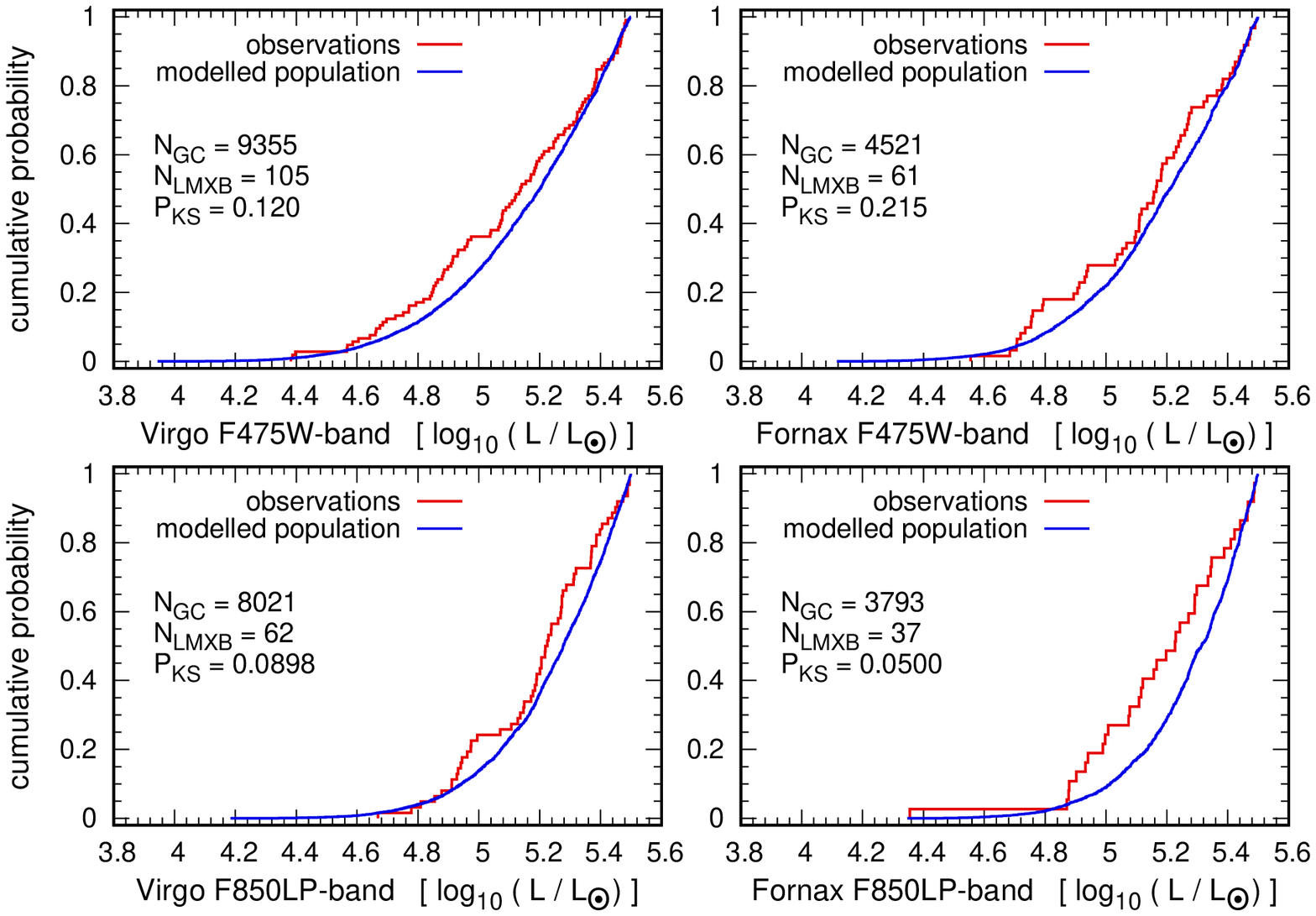}
\caption{\label{fig:l5e5} Like Fig.~\ref{fig:full}, but only for GCs less luminous than $5.5 \times \log_{10}(L / {\rm L}_{\odot})$.}
\end{figure*}

Especially at low luminosities, the observed distributions for the VGC and the FGC have many counts for an LMXB compared to the theory. A reason could be that there are also some LMXBs in GCs, whose probability to exist does not raise with encounter rate (see e.g. \citealt{Canal1976} and \citealt{Kalogera1998} for examples). We call these LMXBs `primordial LMXBs'. They can in principle appear in any GC, but they are most likely to be found in low-luminosity GCs because of their great number and the sparsity of LMXBs created in encounters in them.

We note that the overabundance of LMXBs in low-mass GCs is also apparent in \citet{Sivakoff2007} and \citet{Dabringhausen2012} for the VGC (\citealt{Dabringhausen2012} uses the data of \citealt{Sivakoff2007}, though). This is apparent in both papers by the bin for the GCs with the lowest luminosities, or masses, which does not follow the trend established by the bins with more luminous GCs. Instead, it has a higher fraction of GCs with an LMXB than would be expected.

If we assume that the number of primordial LMXBs is constant per unit mass, while the dynamical formation of LMXBs raises with the encounter rate, then the dynamical formation of LMXBs will eventually take over the primordial formation. The reason is that $\Gamma_{\rm h}$ raises with raising mass of the GCs. It also drops with increasing radii, but the radii of the GCs are, at least at low masses, independent of their mass.

We thus choose, admittedly rather arbitrarily, to separate the full samples into two subsamples at a luminosity of $5.5 \times \log_{10}(L / {\rm L}_{\odot})$.

In the VGC, the subsamples more luminous than $5.5 \times \log_{10}(L / {\rm L}_{\odot})$ contain 1451 GCs in the F475W-passband and 2753 GCs in the F850LP-passband. 76 of them have an LMXB in the F475W-passband and 118 in the F850LP-passband. The subsamples less luminous than $5.5 \times \log_{10}(L / {\rm L}_{\odot})$ contain 9355 GCs in the GCs in the F475W-passband and 8021 GCs in the F850LP-passband. 105 of them contain an LMXB in the F475W-passband and 62 in the F850LP-passband.

In the FGC, 882 GCs in the F475W-passband and 1561 GCs in the F850LP-passband are more luminous than $5.5 \times \log_{10}(L / {\rm L}_{\odot})$. 68 of them have an LMXB in the F475W-passband and 95 in the F850LP-passband. The subsamples less luminous than $5.5 \times \log_{10}(L / {\rm L}_{\odot})$ contain 4521 GCs in the GCs in the F475W-passband and 3793 GCs in the F850LP-passband. 61 of them contain an LMXB in the F475W-passband and 37 in the F850LP-passband.

Note that the difference between the numbers in the F475W-passband and the F850LP-passband is rather large. This is because the limit to admit GC to the samples is set to $5.5 \times \log_{10}(L / {\rm L}_{\odot})$ in both passbands. Thus, there are many GCs that barely reach the criterion in the F850LP-passband, but are just a little bit too dim to reach the criterion in the F475W-passband.

For the GCs in the VGC more luminous than $5.5 \times \log_{10}(L / {\rm L}_{\odot})$, the probability for the difference to be larger than observed is 50 percent in the F850LP-passband and 85 percent in the F475W-passband according to KS-tests, if both samples were randomly drawn from a parent sample.

Thus, the theoretical expectation and observations agree well for the VGC with luminosities $>5.5 \times \log_{10}(L / {\rm L}_{\odot})$ (see left panels of Fig.~\ref{fig:g5e5}). This is because equation~15 in \citet{Sivakoff2007}, or equation~\ref{eq:gamma} here, was fitted to a portion of the same GCs in \citet{Sivakoff2007} to which it is compared by the programme. However, the good fits between theory and observation for the VGC are not self-evident either, because the data with which equation~(\ref{eq:gamma}) is compared is not {\it precisely} the same. \citet{Sivakoff2007} used the ten brightest galaxies in the ACSVCS \citep{Cote2004} and one extra galaxy to search for LMXBs, while we use the galaxies that are within two degrees of M87 and are in the ACSVCS. Moreover, the methods are different. \citet{Sivakoff2007} used a $\chi ^2$-method to establish the exponents in equation~\ref{eq:gamma}, in \citet{Dabringhausen2012} essentially a linear regression was used, and here a KS-test is performed. A difference between a linear regression and a KS-test is that with a KS-test, a single data point that lies off the theory spoils the complete fit, while with a linear regression, the theory can still be saved with many data points that are close to the theory. This is one reason why the KS-test delivers much better agreement between the observation and the theoretical distribution by omitting the low-mass GCs with an LMXB.

Note that the KS-test performs better in the F475W-passband than in the F850LP-passband, even though \citet{Sivakoff2007} used the F850LP-passband for their fit. This could be because of the afore mentioned differences between the data used in \citet{Sivakoff2007} and the sample used here, as well as the different methods. However, It is also possible that it is because the subsample of GCs being more luminous than $5.5 \times \log_{10}(L / {\rm L}_{\odot})$in the F850LP-passband is nearly twice as large than the subsample of GCs being more luminous than $5.5 \times \log_{10}(L / {\rm L}_{\odot})$ in the F475W-passband. Thus, it is difficult to compare the two subsamples directly to each other.

The data from the FGC is however a completely independent sample that is different to the data from the VGC. Hence, there is in principle no reason why the theoretical distribution and the observed data should agree with one another, except there are no differences, or the differences do not matter for the theory (that is, also in the FGC all LMXBs in GCs form by equation~\ref{eq:gamma}, with its exponents fitted for some galaxies in the VGC).

Looking at the data for the FGC for luminosities above $5.5 \times \log_{10}(L / {\rm L}_{\odot})$, there are either too many GCs with luminosities of $5.5 \le \log_{10}(L / {\rm L}_{\odot}) \apprle 5.8$ and an LMXB, or too few GCs with luminosities $\log_{10}(L / {\rm L}_{\odot}) \apprge 5.8$ and an LMXB, to match the theoretical prediction (see right panels of Fig~\ref{fig:g5e5}). Note that the importance of the absolute number of observed GCs is secondary next to their distribution. However, with the observations as they are, the probability for the difference in the FGC to be larger than observed is 1.1 percent in the F850LP-passband and 0.4 percent in the F475W-passband according to a KS-test, if both samples were randomly drawn from a parent sample. Thus, in contrast to the VGC, the comparison of the observations to the theory deteriorates a lot compared to the case with all GCs with an LMXB. Moreover, the fit in the F850LP-passband is better than in the F475W-passband, also in contrast to the VGC.

Note that according to Fig.~\ref{fig:g5e5}, the choice of the GCs in the FGC for distinguishing low-luminosity GCs from high-luminosity GCs could be better still for proving an inconsistency between theory and observations. Instead of choosing $L=5.5 \times \log_{10}(L / {\rm L}_{\odot}) \ (\approx 3 \times 10^5 \ {\rm L}_{\odot})$ for both passbands, Fig.~\ref{fig:g5e5} suggest that $5.6 \times \log_{10}(L / {\rm L}_{\odot}) \ (\approx 4 \times 10^5 \ {\rm L}_{\odot})$ would be a better choice for the F475W-passband, and $L=5.9 \times \log_{10}(L / {\rm L}_{\odot}) \ (\approx 8 \times 10^5 \ {\rm L}_{\odot})$ for the F850LP-passband. It is only above these values that the cumulating curve for the observations rises more steeply than the theoretical curve, while both curves lie pretty much on top of each other below these values. However, the primary reason for choosing  $L=5.5 \times \log_{10}(L / {\rm L}_{\odot})$ as a division was to exclude the range where the primordial LMXBs become dominant in the subsample for the lower luminosities, while keeping the number of GCs with an LMXB still high in the subsample with the higher luminosities.

Given that the reason for separating the full samples into subsamples of high and low luminosities was to keep the pollution from primordial GCs in the subsamples of lower luminosities, is perhaps a surprise that the maximum distances between the theoretical and the observational data are as low as they are. Consequently, the probabilities are high in the subsamples at lower luminosities (Fig.~\ref{fig:l5e5}). This is probably due to how the full sample is divided. This was done such that the theoretical and the observational subsamples at low luminosities both contain the data for GCs with an LMXB and a luminosity below $5.5 \times \log_{10}(L / {\rm L}_{\odot})$, and the theoretical and the observational subsamples at high luminosities only the data above that limit. This is equivalent to forcing the two cumulative functions to intersect at an x-value of $5.5 \times \log_{10}(L / {\rm L}_{\odot})$ in the full samples (Sections~\ref{sec:full-lmxb} and~\ref{sec:full-wo-largest-lmxb}). According to Figs.~\ref{fig:full} and~\ref{fig:full-wo-largest} however, they do not intersect at this luminosity (if ever except at their starting and ending points), unless this additional constraint is made. Without it, the separation of the two cumulative functions can be wider, and the probabilities can therefore be lower. On the other hand, the same argument makes wide separations and small probabilities more meaningful when they do occur, as for the GCs more luminous than $5.5 \times \log_{10}(L / {\rm L}_{\odot})$ with an LMXB in the FGC.

Compared with the data for GCs with LMXBs and luminosities above $5.5 \times \log_{10}(L / {\rm L}_{\odot})$, the KS-tests for all GCs with LMXBs and luminosities below $5.5 \times \log_{10}(L / {\rm L}_{\odot})$ give worse consistencies in the case of the GCs with an LMXB in the VGC, but better consistencies in the case of the FGC according to Fig.~\ref{fig:l5e5}. However, all KS-tests perform better for GCs with an LMXB and luminosities below $5.5 \times \log_{10}(L / {\rm L}_{\odot})$ than for GCs without any restrictions on their luminosity (Sections~\ref{sec:full-lmxb} and~\ref{sec:full-wo-largest-lmxb}). 

\subsubsection{Dividing the LMXB-samples at $5.5 \times \log_{10}(L / {\rm L}_{\odot})$, and without the GCs in the most massive galaxies}
\label{sec:5e5-wo-largest-lmxb}

\begin{figure*}
\centering
\includegraphics[scale=0.95]{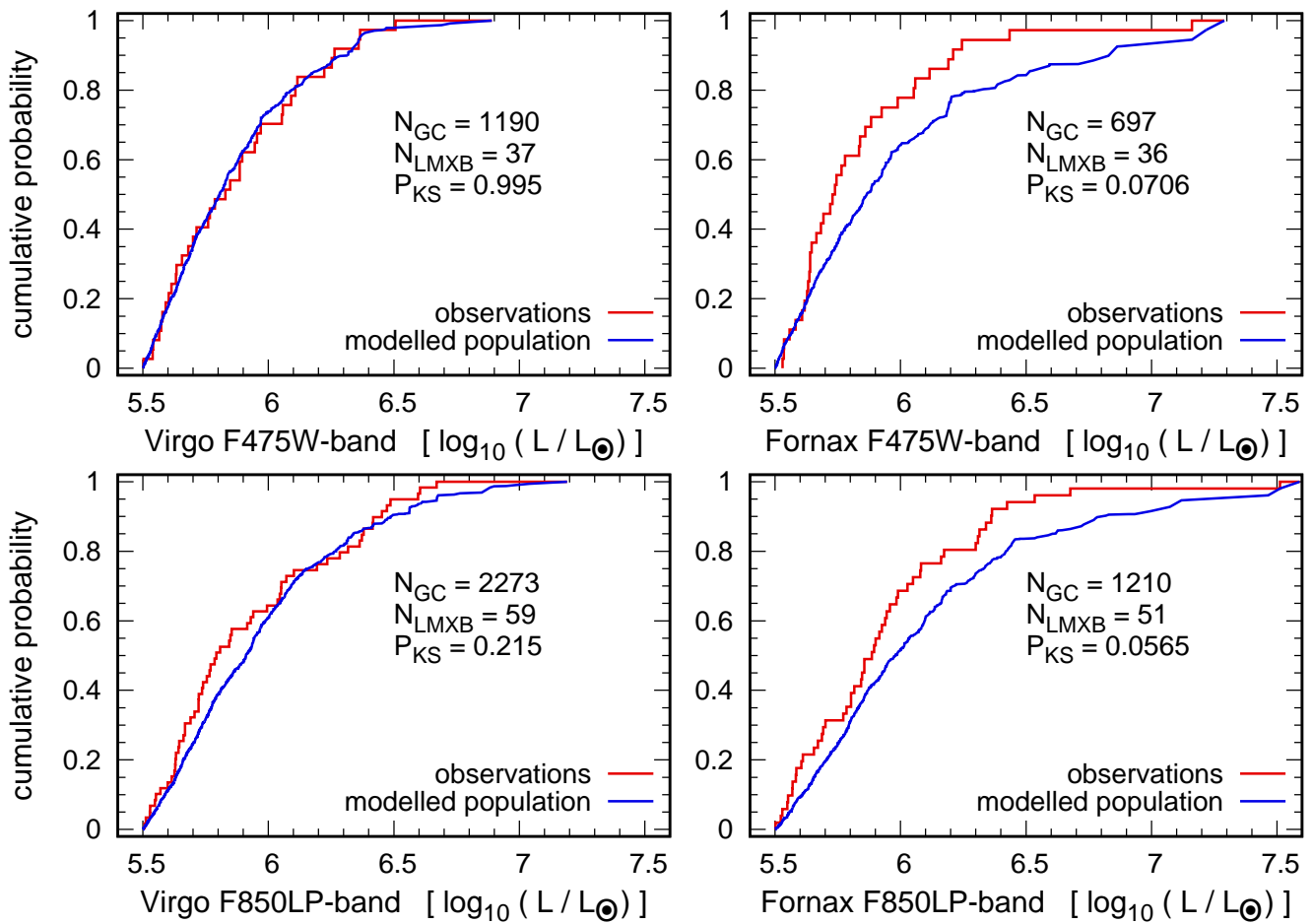}
\caption{\label{fig:g5e5-wo-largest} Like Fig.~\ref{fig:full}, but only for galaxies more luminous than $5.5 \times \log_{10}(L / {\rm L}_{\odot})$. Also, GCs around M87 are excluded for Virgo and GCs around NGC1399 are excluded for Fornax; that is the massive galaxy at the centre is missing for both galaxy clusters.}
\end{figure*}

We once again exclude the GCs of M87 in the VGC and NGC1399 (Section~\ref{sec:full-wo-largest-lmxb}), and combine it with the separation of the full samples into two subsamples at a luminosity of $5.5 \times \log_{10}(L / {\rm L}_{\odot})$ (Section~\ref{sec:5e5-lmxb}).

For the VGC, this leaves 1190 GCs in the F475W-passband and 2273 GCs in the F850LP-passband, which are more luminous than $5.5 \times \log_{10}(L / {\rm L}_{\odot})$. 37 of them have an LMXB in the F475W-passband and 59 in the F850LP-passband. The FGC contains 697 GCs in the F475W-passband and 1210 GCs in the F850LP-passband which are more luminous than $5.5 \times \log_{10}(L / {\rm L}_{\odot})$. 36 of them have an LMXB in the F475W-passband and 51 in the F850LP-passband.

Consequently, the VGC contains 8100 GCs in the F475W-passband and 6989 GCs in the F850LP-passband which are less luminous than $5.5 \times \log_{10}(L / {\rm L}_{\odot})$. 59 of them have an LMXB in the F475W-passband and 37 in the F850LP-passband. The FGC contains 3814 GCs in the F475W-passband and 3245 GCs in the F850LP-passband which are more luminous than $5.5 \times \log_{10}(L / {\rm L}_{\odot})$. 36 of them have an LMXB in the F475W-passband and 23 in the F850LP-passband.

According to a KS-test for GCs more luminous than $5.5 \times \log_{10}(L / {\rm L}_{\odot})$ and with an LMXB in the VGC, the consistency for that the theoretical values and the observations were randomly drawn from the same parent distribution is near 100 percent in the F475W-passband and 22 percent in the F850LP-passband. This is opposed to 85 percent in the F475W-passband and 50 percent in the F850LP-passband when also M87 is included (Fig.~\ref{fig:g5e5}).

Doing the same for the FGC, the consistency for that the theoretical values and the observations of GCs with an LMXB were randomly drawn from the same parent distribution is 7 percent in the F475W-passband and 6 percent in the F850LP-passband. This is opposed to 0.4 percent in the F475W-passband and 1.4 percent in the F850LP-passband when also NGC1399 is included.

In the F475W-passband, this finding is however based on only 11 galaxies in the VGC, and 5 galaxies in the FGC, respectively. In the F850LP-passband, the numbers increase to 14 galaxies in the VGC and to 7 galaxies in the FGC. This means for instance for the F475W-passband that 6 galaxies in the VGC and 4 galaxies in the FGC have only LMXBs in GCs which have luminosities $L<5.5 \times \log_{10}(L / {\rm L}_{\odot}$. On the other hand, these galaxies cannot contain many GCs with an LMXB, because otherwise there would too large a chance that at least one of them would have a higher luminosity than $L=5.5 \times \log_{10}(L / {\rm L}_{\odot}$. This follows from the fact that the division at a luminosity of $L=5.5 \times \log_{10}(L / {\rm L}_{\odot}$ leaves a sizable fraction of GCs with an LMXB at either side of the limit in both passbands. The distribution of the GCs over the galaxies, and thus also the distribution of GCs with an LMXB, is indeed rather uneven. For instance, from the remaining galaxies in the VGC, M89 is responsible for over one half of the GCs with a LMXB. With the FGC it is similar, although the GCs are a bit better distributed over the galaxies.

Nevertheless, for the GCs more luminous than $5.5 \times \log_{10}(L / {\rm L}_{\odot})$, the overall picture stays the same with or without the most massive galaxies. This means good to perfect agreements for in the VGC, but too many GCs with $5.5 \le \log_{10}(L / {\rm L}_{\odot}) \apprle 5.8$, or too few GCs with $\log_{10}(L / {\rm L}_{\odot}) \apprge 5.8$ and an LMXB in the FGC (Fig.~\ref{fig:g5e5-wo-largest}).

However, the GCs in the VGC with an LMXB less luminous than $5.5 \times \log_{10}(L / {\rm L}_{\odot})$ are to 57 percent in the F475W-passband, and to 39 percent in the F850LP-passband, consistent with being drawn from the theoretical sample according to a KS-test. This is opposed to 12 percent and 9 percent for the same kind of tests if M87 is included. Thus, casually speaking, the dim GCs with an LMXB are not consistent with the theoretical distribution if M87 is included, whereas they are if M87 is not. This stands in contrast to the GC with an LMXB less luminous than $5.5 \times \log_{10}(L / {\rm L}_{\odot})$ in the FGC, where the numbers for consistency with being drawn from the theoretical sample are in the same range whether NGC1399 is included or not.

Thus, generally speaking, the results here confirm the results with the largest galaxies included, although with less significance. The exception from this general finding are the GCs less luminous than $5.5 \times \log_{10}(L / {\rm L}_{\odot})$ in the VGC. However, as in the case in Section~\ref{sec:5e5-lmxb}, dividing the theoretical and the observational cumulative functions into two subsamples by the same luminosity limit is equivalent to forcing them to have an intersection at an x-value $5.5 \times \log_{10}(L / {\rm L}_{\odot})$. Thus, the maximum possible distances are lower than in the case without the additional constraint at an x-value of $5.5 \times \log_{10}(L / {\rm L}_{\odot})$, and consequently the probabilities are higher.

\subsubsection{Compact vs. extended globular clusters}
\label{sec:compact}

\begin{figure*}
\centering
\includegraphics[scale=0.95]{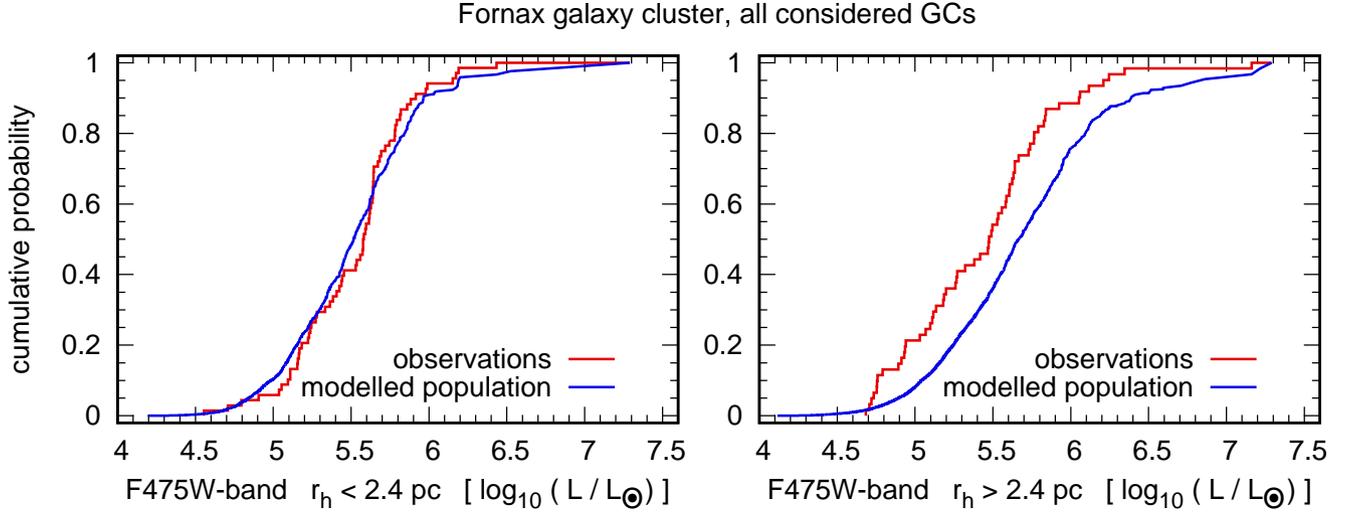}
\caption{\label{fig:full-g-fornax}
Comparisons between the full sample of GCs in the F475W-passband with an LMXB in Fornax and the theoretical distribution for GCs with an LMXB according to equation~15 in \citet{Sivakoff2007}, or equation~\ref{eq:gamma} here. Note that the equation in \citet{Sivakoff2007} was actually fitted to the VGC in the F850LP-passband, so deviations between theory and observations are from that side not completely unexpected. The left panel is for the comparison between the GCs with $r_{\rm h} < 2.4 \, {\rm pc}$ and the right panel for the comparison between the GCs with $r_{\rm h} > 2.4 \, {\rm pc}$. The theoretical samples were cut accordingly, that is they contain either only GCs with $r_{\rm h} < 2.4 \, {\rm pc}$ or GCs with $r_{\rm h} > 2.4 \, {\rm pc}$.}
\end{figure*}

\begin{figure*}
\centering
\includegraphics[scale=0.95]{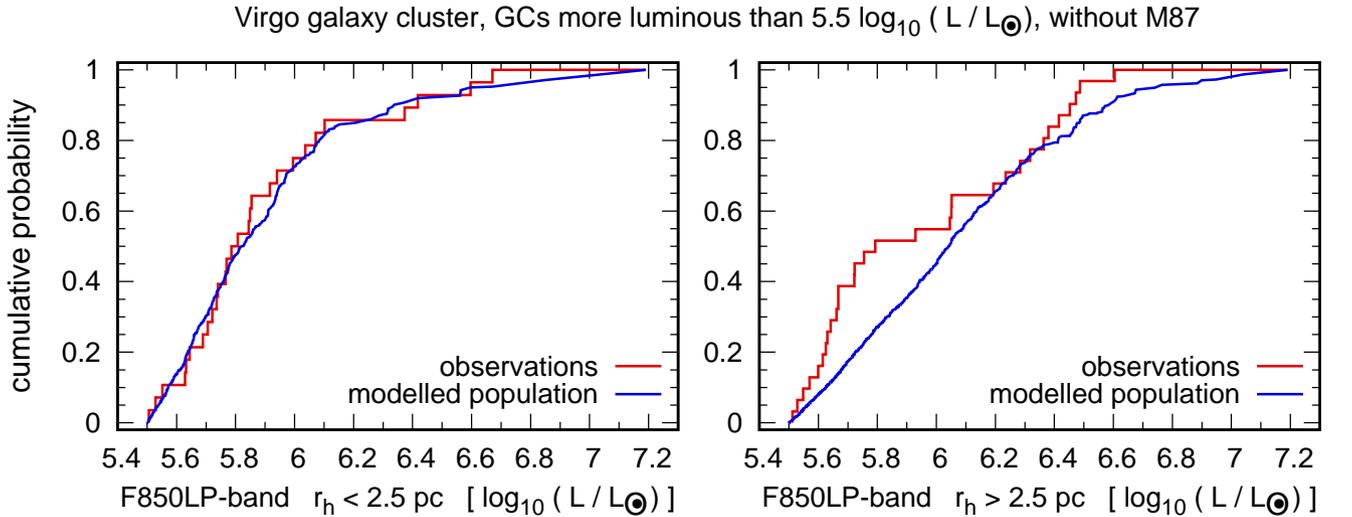}
\caption{\label{fig:5e5-wo-largest-z-virgo}
Like Fig.~\ref{fig:full-g-fornax}, but for the GCs more luminous than $5.5 \times \log_{10}(L / {\rm L}_{\odot})$ in Virgo in the F850LP-passband, and with the GCs around M87 excluded. Note that here the limit between dense and extended GCs is at 2.5 pc and not 2.4 pc, because here GCs in the F850LP-passband are considered, and not in the F475W-passband.}
\end{figure*}

\begin{table*}
\caption{\label{tab:summary} Summary of the the comparisons between the theoretical distributions of the luminosities of GCs with LMXBs and the observations. The headers in the table specify the selection of the GCs for the following four lines. For the theoretical predictions, equation~15 from \citet{Sivakoff2007} was used, or equation~\ref{eq:gamma} here. The full final samples of GCs in the FGC and GCs in the VGC are included. The first column specifies the galaxy cluster and the second column the optical band in which it was observed. The third column tells how many GCs meet the requirements to be included in the final sample for the galaxy cluster in that band, and the fourth column how many GCs have at least one LMXB that is bright enough to be detected. The fifth column gives the probability according to a KS-test that the distance between the theoretical distribution and the observed distribution is even larger than the case realised here, if both distributions are drawn from the same parent distribution. The sixth to the eighth Column are as the third to the fifth column, but only for the GCs with $r_{\rm h}>2.4\, {\rm pc}$ in the F475W-passband and with $r_{\rm h}>2.5\, {\rm pc}$ in the F850LP-passband. The ninth to the eleventh Column are as the third to the fifth column, but only for the GCs with $r_{\rm h} \le 2.4\, {\rm pc}$ in the F475W-passband and with $r_{\rm h} \le 2.5\, {\rm pc}$ in the F850LP-passband.}
\centering
\vspace{2mm}
\begin{tabular}{ll|rrl|rrl|rrl}
\hline
&&&&&&&&&& \\ [-10pt]
Cluster &  band & \multicolumn{3}{|c|}{all radii} &  \multicolumn{3}{|c|}{large radii} & \multicolumn{3}{|c|}{small radii} \\
	       &       & GCs     & LMXBs & prob.  & GCs & LMXBs & prob. & GCs & LMXBs & prob.   \\
\hline
&&&&&&&&&& \\ [-10pt]
\multicolumn{11}{|c|}{all GCs}\\
\hline
Virgo       &   F475W  &  10806  & 181      & 0.0583    &  6792   &  77    &   0.226    & 4014    & 104         & 0.0288  \\     
Virgo       &   F850LP &  10774  & 180 	    & 0.0840    &  7269   &  89    &   0.00500  & 3505    &  91         & 0.247   \\
Fornax      &   F475W  &   5403  & 129      & 0.00450   &  4026   &  61    &   0.00100  & 1377    &  68         & 0.430   \\ 
Fornax      &   F850LP &   5354  & 132      & 0.0136    &  3829   &  66    &   0.132    & 1525    &  66         & 0.230   \\
&&&&&&&&&& \\ [-10pt]
\hline
\multicolumn{11}{|c|}{all GCs except those in the largest galaxy in either galaxy cluster (M87 in the VGC and NGC1399 in the FGC)}\\
\hline
Virgo       &   F475W  &  9290   &  96       & 0.0316   &  5974   &   44        & 0.0951  &  3316   &   52        & 0.0297 \\
Virgo       &   F850LP &  9262   &  96       & 0.0211   &  6295   &   48        & 0.00169 &  2968   &   48        & 0.373  \\ 
Fornax      &   F475W  &  4511   &  72       & 0.0790   &  3470   &   37        & 0.109   &  1041   &   35        & 0.550  \\
Fornax      &   F850LP &  4455   &  74       & 0.0574   &  3279   &   34        & 0.0499  &  1176   &   40        & 0.0731 \\
&&&&&&&&&& \\ [-10pt]
\hline
\multicolumn{11}{|c|}{only GCs $>5.5 \times \log_{10}(L / {\rm L}_{\odot})$}\\
\hline
Virgo       &   F475W  &  1451   &  76       & 0.848    &   965   &   39        & 0.602   &  486   &   37        & 0.882  \\
Virgo       &   F850LP &  2753   & 118       & 0.500    &  1810   &   60        & 0.0746  &  943   &   58        & 0.835  \\ 
Fornax      &   F475W  &   882   &  68       & 0.000381 &   663   &   28        & 0.00507 &  219   &   40        & 0.0503 \\
Fornax      &   F850LP &  1561   &  95       & 0.0114   &  1053   &   48        & 0.300   &  508   &   47        & 0.0149 \\
&&&&&&&&&& \\ [-10pt]
\hline
\multicolumn{11}{|c|}{only GCs $ \le 5.5 \times \log_{10}(L / {\rm L}_{\odot})$}\\
\hline
Virgo       &   F475W  &  9355   & 105       & 0.120    &  5570   &   32        & 0.0933  &  3785  &   73        & 0.0322  \\
Virgo       &   F850LP &  8021   &  62       & 0.0898   &  5175   &   27        & 0.0948  &  2846  &   35        & 0.0578  \\ 
Fornax      &   F475W  &  4521   &  61       & 0.215    &  3226   &   29        & 0.239   &  1295  &   32        & 0.877   \\
Fornax      &   F850LP &  3793   &  37       & 0.0500   &  2540   &   17        & 0.0188  &  1253  &   20        & 0.930   \\
&&&&&&&&&& \\ [-10pt]
\hline
\multicolumn{11}{|c|}{only GCs $>5.5 \times \log_{10}(L / {\rm L}_{\odot})$ except those in the largest galaxy in either galaxy cluster}\\
\hline
Virgo       &   F475W  &   1190  &   37      &  0.995   &   812    &  19         & 0.990    &  378 &     18      & 0.981  \\
Virgo       &   F850LP &   2273  &   59      &  0.215   &  1485    &  31         & 0.0279   &  728 &     28      & 0.904  \\
Fornax      &   F475W  &    697  &   36      &  0.0706  &   554    &  19         & 0.249    &  143 &     17      & 0.115  \\
Fornax      &   F850LP &   1210  &   51      &  0.0565  &   852    &  24         & 0.411    &  358 &     27      & 0.0825 \\
&&&&&&&&&& \\ [-10pt]
\hline
\multicolumn{11}{|c|}{only GCs $\le 5.5 \times \log_{10}(L / {\rm L}_{\odot})$ except those in the largest galaxy in either galaxy cluster}\\
\hline
Virgo       &   F475W  &   8100  &   59      &  0.574   &  4924    &  22         & 0.920    &  3176 &     37      & 0.115  \\
Virgo       &   F850LP &   6989  &   37      &  0.393   &  4585    &  17         & 0.196    &  2416 &     20      & 0.424  \\
Fornax      &   F475W  &   3814  &   36      &  0.0784  &  2815    &  19         & 0.861    &   999 &     17      & 0.0588 \\
Fornax      &   F850LP &   3245  &   23      &  0.2003  &  2229    &  10         & 0.0517   &  1016 &     13      & 0.965  \\
&&&&&&&&&& \\ [-10pt]
\hline
\end{tabular}
\end{table*}

Figs.~\ref{fig:full} to~\ref{fig:g5e5-wo-largest} compare the theoretical distributions with the observed ones as functions of luminosity. However, radii and metallicities do play a decisive role here as well. Smaller radii, higher metallicities or both make the theoretical distributions for the probability for a GC to have a LMXB rise faster, independent of whether these quantities are actually shown or not.

Nevertheless, the influence of the radii is tested further here. We do this by dividing each sample into a subsample which only contains the GCs with compact radii, and a subsample containing the reminder of the GCs.

To set the limit between compact and extended GCs, we searched for the median of the radii of GCs with LMXB in the four samples with $L>5.5 \times \log_{10}(L / {\rm L}_{\odot})$ and without M87, and NGC1399, respectively (see Section~\ref{sec:5e5-wo-largest-lmxb}). In more detail, a specific GC with an LMXB was picked for each such sample. For the FGC in the F850LP-passband, this is for instance the GC that contains the X-ray source 2CXO$\_$J033672.4-345901. These GCs are kept also when the the samples are expanded again, that is when NGC1399 or M87 are included again, or when the GCs less luminous than $5.5 \times \log_{10}(L / {\rm L}_{\odot})$ are considered again. The advantage with choosing the most restricted samples as the starting point is that neither of the two subsamples can die out, because they only grow when the conditions are relaxed again.

Note that for both galaxy clusters, a radius of about 2.4 pc in the F475W-passband, and about 2.5 pc in the F850LP-passband, respectively, was found by this method, although this was not requested. Thus, concerning the peaks of the radius distributions of their bright GCs, the two galaxy clusters without their most massive galaxy appear to be similar.

The detailed results for $N_{\rm GC}$, $N_{\rm LMXB}$ and $P_{\rm KS}$ for GCs with an LMXB and large $r_{\rm h}$, and small $r_{\rm h}$, respectively, can be read out in columns~(6) to~(12) of Table~\ref{tab:summary}. Generally, no real trends can be found. There are some subsamples where the dense GCs fit the theoretical sub-sample while the extended ones do not (GCs with $L>5.5 \times \log_{10}(L / {\rm L}_{\odot})$ in the F850LP-passband in Fornax), some where it is the other way round (all GCs in the F475W-passband in the FGC), some where the observations fit the theoretical sub-samples for both dense and extended GCs (GCs with $L>5.5 \times \log_{10}(L / {\rm L}_{\odot})$ in the F475W-passband in the VGC without M87) and some where both subsamples fail more or less (GCs with $L>5.5 \times \log_{10}(L / {\rm L}_{\odot})$ in the F475W-passband in the FGC).

The argument that the division of the full sample into two subsamples is equivalent to forcing the observational and the theoretical cumulative functions to meet at a specific point in the full sample is however not true here. Here, in contrast to the case in Sections~\ref{sec:5e5-lmxb} and~\ref{sec:5e5-wo-largest-lmxb}, the division into two subsamples is done at a certain radius, while the cumulative functions are still functions of luminosity. The compact GCs as well as the extended GCs can however in principle have any luminosity, while a separation of the GCs into bright and faint GCs (Sections~\ref{sec:5e5-lmxb} and~\ref{sec:5e5-wo-largest-lmxb}) forbids this.

Two especially spectacular examples of comparisons between dense and extended GCs are shown here.

The first one is for all considered GCs in FGC in the F475W-passband (Fig.~\ref{fig:full-g-fornax}). Here, the overabundance of LMXBs in low-luminosity GCs mentioned in Section~\ref{sec:full-lmxb} comes from the extended GCs, while theory and observations fit together nicely in the dense GCs. The trend also exists for the F850LP-passband in Fornax, even though less clearly. In the VGC however, the dense and the extended GCs in both passbands contribute to the over-density of LMXBs for GCs with low luminosities more or less equally.

The second one is for only the GCs more luminous than $5.5 \times \log_{10}(L / {\rm L}_{\odot})$ in Virgo in the F850LP-passband, while GCs around M87 are excluded (Fig.~\ref{fig:5e5-wo-largest-z-virgo}). Here, the extended GCs are overabundant at luminosities slightly above $L>5.5 \times \log_{10}(L / {\rm L}_{\odot})$, while for dense GCs theory and observations fit together nicely for all luminosities. This overabundance of extended GCs with luminosities slightly above $L>5.5 \times \log_{10}(L / {\rm L}_{\odot})$ only happens in F850LP-passband in the VGC. It does not happen in the F475W-passband in the VGC, nor in both considered passbands in the FGC.

Perhaps the main lesson learned from these examples is that the farther the considered sample is away from the sample to which equation~\ref{eq:gamma} was fitted originally, the less predictable a comparison between theory and observations becomes.

\section{Comparison to Dabringhausen et al. (2012)}
\label{sec:Dabringhausen2012}

In \citet{Dabringhausen2012} it was claimed that the frequency of LMXBs in the VGC indicate that the IMF must have been top-heavy in the luminous, and consequently massive GCs, if the LMXBs are formed in stellar encounters. However, in this paper it is claimed based on similar data that the theory of LMXB formation in encounters between stars in GCs fits almost perfectly with the data. There are three reasons for this seeming contradiction.

First, \citet{Dabringhausen2012} compared two samples to each other which are actually different. The first sample in \citet{Dabringhausen2012} contains the the observed GCs with an LMXB in the VGC from \citet{Sivakoff2007}. We call this sample the ´Sivakoff-sample'. The second sample in \citet{Dabringhausen2012} consists of the precise data of 90 UCDs or massive GCs from the VGC, the FGC, Centaurus A and the Local Group. We call this sample the ´UCD-sample'. The reason why the UCD-sample was used in \citet{Dabringhausen2012} despite its small size was that it contains also the velocity dispersions, and thus dynamical masses of the objects, which was essential for their work. The relevant difference here is however that the Sivakoff-sample is based on the data that eventually was published in \citet{Jordan2009} and which contains only objects with $r_{\rm h}<10 \, {\rm pc}$, whereas the UCD-sample contains no restrictions regarding $r_{\rm h}$. This is not a problem in the mass range of ´classical' GCs, which \citet{Mieske2008} defined as GCs with masses below $2\times 10^6 \, {\rm M}_{\odot}$. Almost all of them have indeed $r_{\rm h}<10$ pc. However, above a total mass of $2\times 10^6 \, {\rm M}_{\odot}$, the percentage of objects with half-mass radii $r_{\rm h}>10$ pc increases with increasing total mass\footnote{This is, besides the on average increasing mass-to-light ratios, why they are often considered as a new type of stellar system in the literature, called Ultra-compact dwarf galaxies (UCDs). The radii of the objects in the UCD-sample are shown in fig.~1 in \citet{Dabringhausen2012}.}, and consequently the likelihood to build LMXBs lowers (Section~\ref{sec:LMXBtheory}). Thus, compared to the total number of GCs with masses above $2\times 10^6 \, {\rm M}_{\odot}$ the percentage of GCs with an LMXB will be higher in the Sivakoff-sample than in the UCD-sample, just because of the restriction in $r_{\rm h}$ in the Sivakoff-sample. \citet{Dabringhausen2012} changed the IMF in objects with masses above $2\times 10^6 \, {\rm M}_{\odot}$ in the UCD-sample until the predicted distribution of LMXBs with GC-masses was consistent with the observed LMXBs in the Sivakoff-sample. They thereby found the IMF was top-heavy in such objects, but their results are exaggerated because of the limit to $r_{\rm h}$ in the Sivakoff-sample, which they did not consider.

Second, the way the data in \citet{Dabringhausen2012} is presented could be misleading, because the data is binned. \citet{Dabringhausen2012} plotted the observational data in their fig.~4 at the centres of the bins, which is strictly speaking only appropriate if the data is uniformly, or more generally, symmetrically distributed in each bin. For luminosities resulting into masses above $2\times 10^6 \, {\rm M}_{\odot}$, the observational data is however distributed asymmetrically in such a way, that each bin contains more GCs (or UCDs) that are faint and dense than GCs (or UCD) that are bright and extended. In other words, the GCs at this luminosity range gather more at the faint side of every bin, because they are at the bright side of the peak of the GC luminosity function. In a comparison of the UCD-sample with the observed fraction of GCs with a LMXB (fig.~4 in \citealt{Dabringhausen2012}), the data points showing the observations would ideally be placed at the median of the GC-distribution in each bin instead at the geometrical centre of each bin. At the brightest luminosities, this would put the observational data a bit more to the left in each bin. According to fig.~4 in \citet{Dabringhausen2012}, the observational data continue to rise, while the fractions of GCs with an LMXB in the UCD-sample drop at bright luminosities under the assumption that the IMF is invariant in all considered GCs. The difference between the expectation for an invariant IMF and the observations would therefore appear less drastic if the data points were adjusted to their medians instead of geometrical centres of their bins.

Note however that the IMF of the most massive GCs could be even more top-heavy, if the top-heaviness of the IMF was searched for as in \citet{Dabringhausen2012}, but the observational data was shifted to the left. This is because this transformation would change the luminosities (x-axis) of the data points where the observations are located to lower values, while the fraction of them that has an LMXB in each bin (y-axis) would stay the same. On the other hand, the slope of the fraction of GCs with an LMXB has to be continuous at the transition to the ´classical' GCs (i.e. GCs with masses $<2\times 10^6 \, {\rm M}_{\odot}$). Thus, the slope that has to be achieved by more remnants is steeper still, and thereby works in the opposite direction as excluding objects with $r_{\rm h}> 10 \, {\rm pc}$. 

Finally, and arguably the most important reason are the scalings of the probabilities to have a bright X-ray source in a GC, $P_{X}$. In \citet{Dabringhausen2012}, this is $P_{X}\propto \Gamma_{\rm h} $ or $P_{X}\propto \Gamma_{\rm h}^{0.8} $ and $P_{X}=Z^0=const$, where $\Gamma$ is the encounter rate, and $Z$ is the metallicity of the GC. $\Gamma_{\rm h}$ in turn depends on the mass and the radius of the GC (see Section~\ref{sec:LMXBtheory}). Thus, metallicity effects are neglected in \citet{Dabringhausen2012}, while the exponent to $\Gamma_{\rm h}$ proves to be rather unimportant for the probability to have an LMXB in the GC. \citet{Dabringhausen2012} solved the problem of too little NSs and BHs in massive GCs by making the IMF top-heavy above a luminosity of $10^6 \, {\rm L}_{\odot}$, that is by adding more massive stars. This top-heaviness of the IMF increases with the mass of the GCs. In \citet{Sivakoff2007} however, a dependency on the colour of the GC is considered, too. This colour dependency is interpreted as a dependency on metallicity, which could for instance be because the donor stars would spill over matter for a longer time to the neutron star when they are more metal rich. Thus, metal-rich GCs would be easier to catch in the phase when they have an LMXB. \citet{Sivakoff2007} found $P_{X} \propto \Gamma^{0.82} \, Z^{0.39}$, which is also used here. The IMF is however the same in all considered objects in this scenario. 

Both approaches to set the scalings of the probabilities seem justified at first sight, but according to the available literature, the IMF becoming top-heavy with high masses should definitely play a role in making the fraction of GCs with an LMXB larger. The exponent to the metallicity would then be smaller than 0.39, while zero, as discussed in \citet{Dabringhausen2012}, is perhaps too extreme. This is because other lines of evidence leading also to a change of the IMF with star cluster mass, which leads to more neutron stars than expected. For instance, a change of the integrated galaxy-wide stellar initial mass function (IGIMF) could lead to changes in the mass-to-light ratios of galaxies with galaxy mass (e.g. \citealt{Gunawardhana2011,Fontanot2017,Dabringhausen2019}). The changes of the IGIMF with the star formation rate and and the metallicity of the galaxy is discussed in \citet{Jerabkova2018}, including an overabundance of massive stars with a high star formation rate, like in massive GCs and UCDs. The IGIMF-theory is also able to solve various problems with the chemical abundances of galaxies \citep{Yan2019,Yan2020,Yan2021}.

\section{Summary and conclusion}
\label{sec:summary}

It is well known that the properties of globular cluster systems depend to some extent on the galaxies they accompany. For instance, the GC systems of spiral galaxies are on average smaller than the ones of elliptical galaxies, and the frequency of GCs drops steeper with the distance to the galaxy in low-mass ellipticals than in massive ellipticals (see e.g. \citealt{BrodieStrader2006} and references therein). The question is now how large a system of different galaxies must be for the differences in their GC-systems to level out. Are galaxy clusters with several different galaxies large enough? Or would differences in the galaxy clusters show in their GCs populations instead? For instance, it could be relevant for the GC-systems of whole galaxy {\it clusters} how old those galaxy clusters are, how compact or how large. With this in mind, a comparison of the GC systems of the FGC and the VGC is made in this paper.

For the optical data, the used calalogues of GCs are from \citet{Jordan2007} for the VGC, and from \citet{Jordan2015} for the FGC. They contain both the luminosities and the radii of the GCs. Moreover, the catalogue for the FGC was created after the fashion of the catalogue for the VGC. Additionally, also X-ray point sources were searched for, as indications of low-mass X-ray binaries (LMXBs) in the GCs. This is because the probability for a GC to have an LMXB depends on its luminosity (as an indicator of its mass), its radius and its metallicity. For all X-ray sources, we used the second edition of the Chandra X-ray Observatory source catalogue \citep{Evans2010}. Thus, the optical data and the X-ray data in this paper are as uniform as possible.

We first compared colours, luminosities and radii for the GCs in the VGC with GCs in the FGC with Kologorov-Smirnoff (KS) tests. 

Colours are a quantity that is nearly independent of the distances of the GCs, at least on the distance scales considered here. However, it turns out that the GCs in the FGC are on average a bit bluer than the ones in the VGC. If this was a dependency on redshift, it would be the other way around if anything, because the FGC it a bit farther away than the VGC (20.0 Mpc vs. 16.5 Mpc, \citealt{Blakeslee2009}). Thus, the GCs in the FGC can either be more metal poor, or younger, or both than the GCs in the VGC. It is hard to be more precise, because of a degeneracy of metallicity and age with colour \citep{Worthey1994}. This finding is a manifestation of the well-known fact that more luminous galaxies tend to have redder GC-systems, as the most luminous galaxies of the VGC are more luminous than the ones of the FGC, while no significant difference between both galaxy clusters is visible for the less luminous galaxies. 

In contrast to colours, the apparent radii and luminosities scale directly with the distance of the GCs. If the distances of the galaxy clusters are such that the radii of the GCs in the F850LP-passband are on average the same in both galaxy clusters, then the GCs are on average more compact in the F475W-passband in the VGC. This result depends of course on the ratio of the distance estimates between the two galaxy clusters. For instance, it would also be possible that the GCs have on average the same radii in both galaxy clusters in the F475W band, while in the L850LP-passband, the GCs in Virgo are more extended. Thus, the GCs in the VGC seem a bit more mass segregated on average, because the bluer stars are more massive than the red stars. Therefore the result could be explained if the massive stars in the GCs in the Virgo cluster are more concentrated to the centres of their GCs. However, that the slopes are different between the two band remains, independent of the exact distance ratio. They are nearly parallel in the F850LP-passband, but in the F475W-passband, the slope in the FGC is flatter than in the VGC.

Note that all these differences are found from more than $10^5$ GCs distributed over 100 galaxies in the VGC and 44 galaxies in the FGC. This makes them statistically robust.

We also find about 180 GCs with a bright X-ray source in the VGC and about 130 GCs with a bright X-ray source in the FGC, which we interpret as LMXBs. They are distributed over only 18 galaxies in the VGC and 10 galaxies in the FGC; partly because not every GC in the optical harbours a bright X-ray source, but partly also because we searched only in the two degrees centered on the respective most massive galaxies of the VGC and the FGC. In total, the two degrees around M87 contain 35 of the 100 galaxies covered in the ACSVCS \citep{Cote2004}, and the two degrees around NGC1399 contain 12 of the 44 galaxies covered by the ACSFCS \citep{Jordan2007}. 

The theoretical distribution for the probability of a GC to have a LMXB can be expressed by exponents to the encounter rates and the metallicities of the GCs. The encounter rate is in turn a function of the radius and the mass of the GCs, or luminosity as luminosity is a proxy for mass. Thus, in the end, the probability for a GC to have an LMXB is a function of the masses, the radii and the metallicities of the GCs.

The matches between observations and theory were tested with KS-tests. If the full samples are considered, the probability for the VGC that there is no difference between the sample from which the theoretical data is drawn and the sample from which the observational data is drawn is 8 percent for the F475W-passband and 6 percent for the F850LP-passband. For the FGC, it is about 1 percent for the full samples in both passbands. This mediocre result is because of too many faint, or light, GCs having an LMXB. A possible reason could be that in the observational data, there is an additional channel for the creation of LMXBs, that is independent on encounter rates and metallicities. In the theoretical samples, such alternatives for the creation of LMXBs were however not considered.

If this is true, then cutting off the many low-luminosity GCs, or low-mass GCs, would bring more agreement between theory and practice. This is because the encounter rates become more dominant for the production of LMXBs with the mass of the GCs. Indeed, leaving the GCs with small luminosities aside leads to a good agreement between theory and practice in the VGC. This is no surprise, since almost the same data on the VGC was used as the basis for the theoretical model. However, in the FGC, the probability from the KS-test does not change, or even becomes worse if the same luminosity cutoff like in the Virgo galaxy cluster is used. 

The most massive galaxy in either galaxy cluster is not the reason for the difference in their GC systems, because the general picture prevails when they are left out. However, the results become somewhat less significant.

Thus, there are several indications that the overall GC system in the FGC is indeed different from the one in the VGC. These are differences in the luminosities, radii, and colours of the GCs, and differences in the distributions of their LMXBs.

\section*{Acknowledgements}
J\"{o}rg Dabringhausen gratefully acknowledges support from the Grant Agency of the Czech Republic under grant number 20-21855S, as well as past support from the post-doctoral programme of FONDECYT under grant number 3140146 at the Universidad de Concepci\'{o}n. He also acknowledges two working stays with Steffen Mieske in 2010 and 2015 funded by ESO Chile, and one working stay with Michael Fellhauer in 2019 at the Universidad de Concepci\'{o}n funded by FONDECYT regular No. 1180291. Especially his work in Chile proved valuable for this project. Michael Fellhauer gratefully acknowledges funding through FONDECYT regular No. 1180291 and through the ANID BASAL projects ACE210002 and FB210003. This research has made use of data obtained from the Chandra Source Catalog, provided by the Chandra X-ray Center (CXC) as part of the Chandra Data Archive. We acknowledge the usage of the HyperLeda database (http://leda.univ-lyon1.fr)

\section*{Data availability}
All used data is available in the cited literature. No new data was introduced.

\bibliographystyle{mn2e}
\bibliography{differ-gc}

\label{lastpage}

\end{document}